\documentclass[useAMS,usenatbib]{mn2e}  
\usepackage{graphicx,amssymb,afterpage}
\def\aj{AJ}%
\def\apj{ApJ}%
\def\apjl{ApJ}%
\def\apjs{ApJS}%
\def\aap{A\&A}%
\def\mnras{MNRAS}%
\def\pasj{PASJ}%
\def\sovast{Soviet~Ast.}%
\begin{document}
   \title[Polarization signatures of unresolved radio sources]{Polarization signatures of unresolved radio sources
}

   \author[Schnitzeler, Banfield, \& Lee]
          {D.H.F.M. Schnitzeler$^{1}$\thanks{Schnitzeler@mpifr-bonn.mpg.de}, 
           J.K. Banfield$^{2,3}$
           K.J. Lee$^{4,1}$\\ 
          $^1$ Max Planck Institut f\"ur Radioastronomie, D-53121 Bonn, Germany\\
          $^2$ Australia Telescope National Facility, CSIRO Astronomy and Space Science, Marsfield, NSW 2122, Australia\\
          $^3$ Research school of Astronomy and Astrophysics, Australian National University, Weston Creek, ACT 2611, Australia\\
          $^4$ Kavli Institute for Astronomy and Astrophysics, Peking University, Beijing 100871, Peoples' Republic of China\\
          }

   \date{Accepted 2015 March 27.  Received 2015 March 25; in original form 2014 November 18}

   \pagerange{}\pubyear{}

   \maketitle

 
\begin{abstract}
We investigate how the imprint of Faraday rotation on radio spectra can be used to determine the geometry of radio sources and the strength and structure of the surrounding magnetic fields. 
We model spectra of Stokes $Q$ and $U$ for frequencies between 200 MHz and 10 GHz for Faraday screens with large-scale or small-scale magnetic fields external to the source.
These sources can be uniform or 2D Gaussians on the sky with transverse linear gradients in rotation measure (RM), or cylinders or spheroids with an azimuthal magnetic field.
At high frequencies the spectra of all these models can be approximated by the spectrum of a Gaussian source; this is independent of whether the magnetic field is large-scale or small-scale. 
A sinc spectrum in polarized flux density is not a unique signature of a volume where synchrotron emission and Faraday rotation are mixed. 
A turbulent Faraday screen with a large field coherence length produces a spectrum which is similar to the spectrum of a partial coverage model.
At low and intermediate frequencies, such a Faraday screen produces a significantly higher polarized signal than Burn's depolarization model, as shown by a random walk model of the polarization vectors.
We calculate RM spectra for four frequency windows.
Sources are strongly depolarized at low frequencies, but RMs can be determined accurately if the sensitivity of the observations is sufficient.
Finally, we show that RM spectra can be used to differentiate between turbulent foreground models and partial coverage models.
\end{abstract}
\begin{keywords} magnetic fields -- polarization -- galaxies: ISM -- galaxies: jets -- galaxies: magnetic fields
\end{keywords}

%

\section{Introduction}\label{s-introduction}
Most Active Galactic Nuclei (AGN) are so distant that we can only study their structure using Very Long Baseline Interferometry (VLBI). However, the magnetic field in the medium surrounding AGN leaves an imprint on the polarized flux density spectra, allowing us to study AGN using polarization measurements. 
In a magnetized plasma the refractive index for the left-handed and right-handed circular polarization modes is slightly different, which leads to a net rotation of the linear polarization vector between frequencies, also known as Faraday rotation. 
The amount of Faraday rotation of a polarized wave which is detected at a wavelength $\lambda$ is equal to $\chi_\mathrm{obs} - \chi_\mathrm{0}\ =\ \mathrm{RM}\lambda^2$, where $\chi_\mathrm{obs}$ and $\chi_\mathrm{0}$ indicate the position angle of the polarized electromagnetic wave that is detected and the position angle of the wave when it is emitted, respectively.
The rotation measure RM is equal to
\begin{eqnarray}
\mathrm{RM}\ =\ 0.81 \int_\mathrm{source}^\mathrm{observer} n_\mathrm{e} B_\| \mathrm{d}l\, \mathrm{rad~m}^{-2}\, ,
\label{rm_definition}
\end{eqnarray}
where $n_\mathrm{e}$ is the local electron density in units of cm$^{-3}$, $B_\|$ the line-of-sight component of the magnetic field in units of $\mu\mathrm{G}$, and the path length is measured in parsec.
If $n_\mathrm{e}$, $B_\|$, or the path length changes inside or across the source then equation~\ref{rm_definition} applies to each individual region where these quantities are the same. 
In that case the net observed polarization vector is a linear combination of the polarization vectors which are emitted by volumes with the same RM, and the \emph{net} RM of the source $\equiv \mathrm{d}\chi_\mathrm{obs}/\mathrm{d}\lambda^2$, which can depend on frequency.

Depending on the geometry of the magnetic field, its strength, coherence length, and the electron density distribution along the line of sight, different parts of the background source will have different RMs. 
These regions with different RMs leave an imprint on the polarized flux density and position angle spectrum of the source, which can be decomposed into emission at different RMs using the technique of rotation measure synthesis \citep{brentjens2005}.
In this Paper we investigate how polarization measurements can be used to study the properties of synchrotron-emitting and Faraday-rotating media in unresolved sources.
These sources can be distant AGN or structures in nearby galaxies which are below the resolution limit of a telescope.
We develop and analyse polarized flux density and position angle spectra as a function of frequency, and RM spectra, for different geometries of the sources themselves and of the magnetic fields in their surroundings. 
Analytical models of structured magnetic fields (either large-scale or small-scale) have been presented previously by, e.g., \citet{burn1966}, \citet{sazonov1973}, \citet{jones1977}, \citet{cioffi1980}, \citet{laing1981}, \citet{bicknell1990}, \citet{tribble1991}, and \citet{sokoloff1998}. At the time, polarimeters on radio telescopes only provided data for very narrow frequency windows. To study the structure of magnetic fields in AGN one would typically analyse data at a number of frequencies, as was done for example by \citet{rossetti2008}, \citet{mantovani2009}, and more recently by \citet{farnes2014a}, \citet{farnes2014b}.
Modern radio interferometers like the Jansky Very Large Array and the Australia Telescope Compact Array that are equipped with broad-band polarimeters can be turned into powerful tools for studying the complex structure of AGN and their associated magnetic fields. \citet{farnsworth2011}, and \citet{osullivan2012} studied compact radio sources using broad-band data and the technique of RM synthesis. Building on the work by \citet{burn1966} and \citet{tribble1991}, \citet{arshakian2011} and \citet{bernet2012} have modelled polarized flux density spectra for media that contain large-scale and small-scale magnetic fields. 
Very recently, \citet{horellou2014} presented analytical models of helical magnetic fields and calculated polarized flux density spectra from these.
\citet{shneider2014} developed analytical models of depolarization by media with large-scale or turbulent magnetic fields and used these to interpret their observations of the grand-design spiral galaxy M51. 

In our Paper we consider radio sources which are compact compared to the size of the telescope beam, so that beam attenuation is not important. Faraday rotation occurs in front of the source, and can be due to either large-scale or turbulent magnetic fields.
The large-scale magnetic fields that we consider produce a transverse linear gradient in RM across a uniformly emitting source or a source with a Gaussian flux density profile, or they wrap around one of the axes of the source. 
We consider turbulent magnetic fields that can be characterised by a single coherence length.
We develop semi-analytical models for these different source types, which we use to calculate polarized flux density and position angle spectra as a function of frequency, and RM spectra.
Then we analyse which features in the frequency spectra or RM spectra can be used to identify the geometry of the emitting region and the properties of the Faraday-rotating screen in front of this region.
In our models we will consider RMs of up to at least 1000~rad~m$^{-2}$ to reflect the range in RMs that have been found in AGN cores, see, e.g., \citet{zavala2003, zavala2004}, \citet{osullivan2011}, and \citet{hovatta2012}.
Compared to previous studies, we consider a wider range of models and develop new models, present model spectra for a wider frequency range, include depolarization across the finite frequency channels, and analyse these spectra using RM synthesis. 
The wide frequency range that we consider will be accessible with the Square Kilometre Array (SKA) which is expected to produce its first results around 2020. 

In Section~\ref{s-models} we describe the models that we developed that include large-scale or turbulent (small-scale) magnetic fields, and we present polarized flux density spectra for these source geometries. 
In Section~\ref{sim-rm-spec} we calculate RM spectra from the Stokes $Q$ and $U$ frequency spectra that we modelled for four observing windows. 
In the Appendix we describe in detail how we modelled emitting cylinders and emitting ellipsoids with a wrapped-around Faraday-rotating layer, and we derive probability density functions of RM for turbulent Faraday screens.

\section{Model descriptions and frequency spectra}\label{s-models}
When we integrate through the synchrotron-emitting medium to calculate the polarized flux density profile of the source we make the following assumptions:
\begin{enumerate}
\item The monochromatic volume emissivity $\epsilon_\nu$ and the shape of the emission spectrum are the same throughout the source,
\item The emission is synchrotron-thin; synchrotron self-absorption is not important,
\item The intrinsic position angle of the polarized emission is constant throughout the source, 
\item Wavelength-independent depolarization is constant throughout the source, and
\item The bulk velocity of the synchrotron-emitting plasma is much smaller than the speed of light.
\end{enumerate}
The polarization properties of jets with relativistic bulk velocities have been modelled by, e.g., \citet{blandford1979}, \citet{beckert2002}, \citet{pariev2003}, \citet{lyutikov2005}, \citet{zakamska2008}, 
\citet{broderick2010}, \citet{clausen-brown2011}, and \citet{porth2011}.
These authors have demonstrated that relativistic jets with helical magnetic fields inside the source region show complex polarization behaviour, for example, the electric vector position angle can change direction by 90 degrees between the axis and the edge of the jet (`spine-sheath' structures: see, e.g., \citealt{attridge1999}, \citealt{pushkarev2005}), and the polarized flux density does not have to be symmetric along a cross-cut perpendicular to the axis of the jet.
In our Paper we limit ourselves to sources (which could be jets) with non-relativistic bulk velocities to reduce the complexity of the models, thereby simplifying our analysis.
In all our models the Faraday-rotating gas is non-relativistic; for Faraday rotation in a relativistic gas we refer the reader to the paper by \citet{broderick2009}. 

We will use the `intrinsic polarization fraction' $p_0$ to refer to the polarization fraction that is measured by a hypothetical observer between the source of the emission and the Faraday-rotating foreground screen, while we ourselves measure the polarized emission only after it has passed through the foreground screen. This intrinsic polarization fraction includes wavelength-independent depolarization effects that are not caused by Faraday rotation. 
Dividing the surface-integrated monochromatic polarization vector by $p_0$ times the Stokes $I$ spectrum removes spectral index effects.
We choose the orientation of the Stokes $Q,U$ coordinate system such that the source emits only in Stokes $Q$ (before the emission passes through the Faraday-rotating foreground medium).

The observed monochromatic polarization vector at frequency $\nu$ is given by
\begin{eqnarray}
\bmath{P}_\mathrm{obs}\left(\nu\right) & \equiv &  
\int_\mathrm{source} \bmath{P}_\mathrm{em}\left(x,y,z\right)\mathrm{e}^{2\mathrm{i}\mathrm{RM}\left(x,y,z\right)\left(\mathrm{c}/\nu\right)^2}\mathrm{d}x\mathrm{d}y\mathrm{d}z
\label{p_monochromatic_source} \\
& = & \int_{-\infty}^\infty \bmath{P}_\mathrm{em}\left(\mathrm{RM}\right)\mathrm{e}^{2\mathrm{iRM}\left(\mathrm{c}/\nu\right)^2}\mathrm{dRM}\, , 
\label{p_monochromatic}
\end{eqnarray}
where $\bmath{P}_\mathrm{em}\left(x,y,z\right)$ indicates the polarization vector that is emitted at position $\left(x,y,z\right)$ within the source, $\bmath{P}_\mathrm{em}\left(\mathrm{RM}\right)$ indicates the polarization vector that is emitted at a single RM, and `c' the speed of light.
Throughout this Paper we will use a Cartesian coordinate system $x,y,z$ where $x$ points towards the observer, and $y$ and $z$ lie in the plane of the sky such that $z$ points along the major axis of the source and $y$ along the minor axis. The major axis of the source can be inclined by an angle $\theta$ with respect to the plane of the sky; in that case $y$ and $z$ point along the projection of the minor and the major axis of the source on the sky, respectively.

We will use analytical expressions or numerical modelling to determine $\bmath{P}_\mathrm{obs}\left(\nu\right)$.
Then we calculate the net polarization vector of each frequency channel by integrating over the response function of that channel. For channel $j$
\begin{eqnarray}
\bmath{P}_\mathrm{obs}\left(\nu_j\right) & \equiv & \int_{-\infty}^{\infty} w_j\left(\nu\right)\bmath{P}_\mathrm{obs}\left(\nu\right)\mathrm{d}\nu\, .
\label{channel_RM_depol}
\end{eqnarray}
The response function of channel $j$, $w_j\left(\nu\right)$, expresses how sensitive the observations are to each frequency in the observing band. 
The integral of $w_j\left(\nu\right)$ out to $\pm$ infinity is equal to one.
We will simulate frequency channels with a top-hat response in frequency, which is equal to 1/$\delta\nu$ inside the channel and zero outside it, where $\delta\nu =  \nu_{j+1} - \nu_j$ is the channel width.
We will refer to the channel-averaged polarization vectors as $\bmath{P}_\mathrm{obs}\left(\nu; \delta\nu\right)$.
The polarized flux density that the background source emits is independent of the structure of the Faraday screen, therefore
\begin{eqnarray}
\int_{-\infty}^{\infty}\left| \bmath{P}_\mathrm{em}\left(\mathrm{RM}\right)\right|\mathrm{dRM}\ =\ \mathrm{constant}\, .
\label{sum_P_constant}
\end{eqnarray}
In our simulations the background source emits 1000 units of polarized flux density. 
The physical interpretation of equation~\ref{sum_P_constant} is that structure in the Faraday screen redistributes polarized flux density over different RMs. 
As a result sources that emit over a wider range of RMs have a lower peak polarized flux density in the RM spectrum.

\subsection{A foreground medium with a large-scale magnetic field}\label{channel_depolarization_structured_screen}
In this section we simulate transverse linear RM gradients in front of a source that emits uniformly across its surface (\S~\ref{burnjet.sec}) or has a Gaussian emission profile (\S~\ref{gaussjet.sec}). Such RM gradients can occur when the source of the emission is embedded in a large ionized halo. If the source is small compared to the radius of the halo then any inclined jet will show a transverse linear RM gradient across its surface because of the increase in path length through the halo.
Next we model sources where the geometry of the Faraday-rotating foreground and the geometry of the emitting region are closely connected. In \S~\ref{cylinder.sec} we consider a magnetic field that wraps around a cylinder which emits polarized radio waves; the magnetic field inside the emitting cylinder points along the major axis of the cylinder.
In \S~\ref{ellipsoid.sec} we generalize the shape of the emitter to an ellipsoid that can have any axis ratio and any orientation with respect to the line of sight. 
Azimuthal magnetic fields have been observed in the radio jets of 3C273 (\citealt{asada2002}), M87 (\citealt{algaba2013}), and in other radio jets (e.g., \citealt{reichstein2012}, \citealt{gabuzda2014}), 
even though in those cases the synchrotron-emitting plasma moves down the jet at relativistic speeds, while for the cylinders and ellipsoids that we model we consider plasmas with non-relativistic bulk velocities.
The expressions we derive are valid for a single source; if a source consists of multiple components  then the resulting RM spectrum is simply the vector sum of the complex RM spectra of the individual components. 

To avoid unnecessary calculations, when we use numerical integration we increase the \emph{(fractional)} numerical accuracy of the integral at the highest frequencies, where the integrated polarized flux densities are highest. This guarantees that the integrated flux densities at all frequencies are accurate down to at least $10^{-5}$ flux density units, and that we do not spend too much time numerically integrating at the lowest frequencies. We start our simulation always at the highest frequency, where depolarization is minimal, and for this frequency we set the fractional numerical accuracy $\epsilon$ = 10$^{-10}$. Then we move to lower frequencies, and we let $\epsilon$ for channel $j$ increase as 
\begin{eqnarray}
\epsilon_{j+1}\ =\ 10^{-\mathrm{floor}[\log_{10}\left(\mathrm{abs}\left(\bmath{P}_{\mathrm{obs},j}\right)/0.01\right)+5]}\, ,
\label{epsilon_definition}
\end{eqnarray}
up to 10$^{-5}$. When numerically integrating non-monotonic functions we checked that the numerical integral was calculated with a sufficiently small (accurate) $\epsilon$.
In \S~\ref{ellipsoid.sec} we will use a different method to fix the numerical accuracy with which polarization vectors are calculated for each of the frequency channels.
The accuracy of our numerical method is sufficient to calculate RM spectra in Section~\ref{sim-rm-spec} down to at least 0.01 flux density units (-50 dB).

\subsubsection{Uniform source with a transverse linear RM gradient}\label{burnjet.sec}
The net monochromatic polarization vector $\bmath{P}_\mathrm{obs}\left(\nu\right)$ for a uniform source of emission inside which RM increases or decreases linearly with physical depth has been calculated analytically by \citet{burn1966}:
\begin{eqnarray}
\bmath{P}_\mathrm{obs}\left(\nu\right)\ & = & \left( p_0\times I\right)\mathrm{sinc}\left(\Delta\mathrm{RM} \left(\mathrm{c}/\nu\right)^2\right) \mathrm{e}^{2\mathrm{i}\mathrm{RM_c}\left(\mathrm{c}/\nu\right)^2}\, ,
\label{channel_rm_depol_simplified}
\end{eqnarray}
where $\Delta\mathrm{RM}\ =\ \mathrm{RM_\mathrm{max}}\ -\ \mathrm{RM_\mathrm{min}}$ is the range in RM over which the source emits, and $\mathrm{RM_c}\ =\ 0.5\mathrm{RM_\mathrm{min}}\ +\ 0.5\mathrm{RM_\mathrm{max}}$ is the mean RM of the emission. 
We define $\mathrm{sinc}\left(x\right) \equiv \mathrm{sin}\left(x\right)/x$. 
Such a source in which emission and Faraday rotation are mixed has become known as a `Burn slab'.
Equation~\ref{channel_rm_depol_simplified} also describes the spectrum of a uniformly emitting source which has a linear RM gradient in front of it (E.g., \citealt{sokoloff1998}, and \citealt{zavala2004}).
Fig.~\ref{burn_jet_spectra.fig} shows the absolute value of the simulated polarized flux density $\bmath{P}_\mathrm{obs}\left(\nu; \delta\nu\right)$, scaled with the assumed polarized fraction times the flux density spectrum in Stokes $I$, of uniform sources that lie behind linear gradients in RM. We integrated  $\bmath{P}_\mathrm{obs}\left(\nu\right)$ across frequency channels using Romberg's method. 

Equation~\ref{channel_rm_depol_simplified} expresses how the emission at different RMs leads to depolarization (the sinc term) while the polarization vectors at different frequencies show Faraday rotation with a \emph{net} RM equal to the mean RM of the emission.
If $\mathrm{RM_c}$ = 0 rad~m$^{-2}$, then the observed polarization vectors do not show net Faraday rotation, and all polarized emission will be in Stokes $Q$, Stokes $U$ will be 0. 

The condition that $\left|\bmath{P}_\mathrm{em}\left(\mathrm{RM}\right)\right|$ is the same for all RM is much more general than one might think at first. 
There are no constraints on the geometry of the source, nor on the inclination of the source with respect to the plane of the sky. 
The source can even consist of multiple components.
The RMs across the source do not have to increase or decrease monotonically, and the RM gradient can have any orientation with respect to the major axis of the background source.
For example, a cylindrical source with a linear RM gradient along its major axis is described by equation~\ref{channel_rm_depol_simplified}.
Even the spectrum of a source with an azimuthal magnetic field can be approximated by equation~\ref{channel_rm_depol_simplified} if the emission profile along the RM gradient of the source is approximately constant. 
Unfortunately, because all these geometries have the same frequency and RM spectra it is not possible to determine the source geometry without additional information.

\subsubsection{Gaussian source with a transverse linear RM gradient}\label{gaussjet.sec}
\citet{leahy1986}, \citet{johnson1995}, and \citet{sokoloff1998} calculated monochromatic polarization vectors for a uniformly emitting source with a linear gradient in RM in front of it, which is observed with a Gaussian beam.
We note that by re-ordering the factors inside the integrals that these authors solved the same solution is found for a source with a 2D Gaussian profile that lies behind a Faraday screen with a transverse linear gradient in RM.
The Gaussian source measures $\sigma_y\times\sigma_z$ standard deviations along its minor and major axis, respectively, and the RM gradient is described by the function RM$\left(y,z\right)$.
The observed polarization vector at a single frequency is then given by equation~40 in \citet{sokoloff1998}:
\begin{eqnarray}
\bmath{P}_\mathrm{obs}\left(\nu\right) & = &
\frac{\left( p_0\times I\right)}{2\upi\sigma_y\sigma_z}  
\int_{-\infty}^{\infty}\int_{-\infty}^{\infty} 
\mathrm{e}^{-\frac{1}{2}\left[\left(\left(y-y_0\right)/\sigma_y\right)^2+\left(\left(z-z_0\right)/\sigma_z\right)^2\right]}
\nonumber\\
& & \mathrm{e}^{2\mathrm{iRM}\left(y,z\right)\left(\mathrm{c}\right/\nu)^2}\mathrm{d}y\mathrm{d}z\, \nonumber\\
& = & \left( p_0\times I\right) \mathrm{e}^{-\ 2\Delta\mathrm{RM}^2\left(\mathrm{c}/\nu\right)^4\ +\ 2\mathrm{i}\mathrm{RM_c}\left(\mathrm{c}/\nu\right)^2}\, , 
\label{rm_gradient_plus_gaussian}
\end{eqnarray}
where 
\begin{eqnarray}
\Delta\mathrm{RM}^2 & = & \left(\frac{\partial\mathrm{RM}}{\partial  y}\sigma_y\right)^2 + \left(\frac{\partial\mathrm{RM}}{\partial  z}\sigma_z\right)^2 \nonumber
\end{eqnarray}
is the total change in RM across the Gaussian source, and `RM$_\mathrm{c}$' is the RM at the centre of the Gaussian source on the sky, at coordinates $\left(y_0,z_0\right)$. 
$\Delta\mathrm{RM}^2$ can be expressed in terms of the RM difference across the FWHM of the major and minor axes of the source by using $\sigma = \mathrm{FWHM}/\sqrt{8\mathrm{ln}2}$.

We numerically integrate equation~\ref{rm_gradient_plus_gaussian} over the widths of the individual frequency channels using Romberg integration. Because $\left| \bmath{P}_\mathrm{obs}\left(\nu\right)\right|$ decreases monotonically with decreasing $\nu$, and because the channel width $\delta\nu$ is constant, $\left|\bmath{P}_\mathrm{obs}\left(\nu; \delta\nu\right)\right|$ decreases monotonically with decreasing $\nu$. Starting at the highest observing frequencies, we stopped integrating equation~\ref{rm_gradient_plus_gaussian} at the frequency where  $\left|\bmath{P}_\mathrm{obs}\left(\nu; \delta\nu\right)\right|/\left(p_0\times I\right)$ was $<$ 10$^{-5}$, which is below the limit we consider in Section~\ref{sim-rm-spec} for calculating RM spectra.

\begin{figure}
\resizebox{\hsize}{!}{\includegraphics{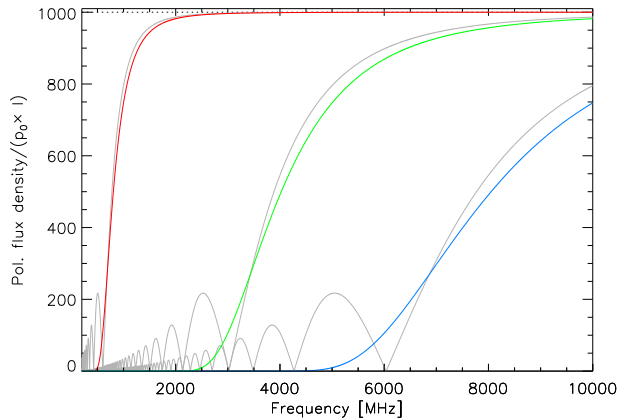}}
\caption{
Broad-band polarized flux density spectra between 200 MHz and 10 GHz, scaled with the intrinsic polarization fraction times the Stokes $I$ spectrum, for Gaussian background sources that illuminate a linear gradient in RM. The source emits 1000 flux density units. The RM differences across the FWHM of the major axis of the Gaussian are 10, 250, and 1000 rad~m$^{-2}$ (red to blue). 
The grey lines in the background show spectra for uniform sources from \S~\ref{burnjet.sec} with a half-width equal to three times the standard deviation of each of the simulated Gaussian sources. These uniform sources emit over RM ranges of 13, 318, and 1274 rad~m$^{-2}$. The spectrum of a Gaussian source and the matching spectrum of a uniform source intersect at approximately 300 flux density units. 
}
\label{burn_jet_spectra.fig}
\label{gauss_jet_spectra.fig}
\end{figure}

Fig.~\ref{gauss_jet_spectra.fig} shows frequency spectra of Gaussian sources with transverse linear RM gradients between 10 and 1000 rad~m$^{-2}$ FWHM$^{-1}$.
Gaussian emitters with even a small RM difference across the source ($\gtrsim$ 125 rad~m$^{-2}$~FWHM$^{-1}$) are almost completely depolarized at frequencies below 1 GHz.
This is not the result of depolarization across the frequency channels, which we checked by simulating observations with frequency channels of 0.1 MHz.
When the RM gradient is small, the polarized flux density spectrum shows a sharp drop over a narrow frequency range. For such sources it is vital to match the frequency window of the observations to the RM gradient one is looking for.
If the observing frequency is too low the source will be severely depolarized and therefore difficult to detect. On the other hand, if the observing frequency is too high the source will not show any depolarization, and will be indistinguishable from other sources with small RM gradients which also do not show depolarization.

If a source has a Gaussian profile along its minor axis and a uniform profile along its major axis, and illuminates a linear RM gradient, then the orientation of the RM gradient determines if the spectrum we observe follows the sinc profile from \S~\ref{burnjet.sec} (RM gradient along the major axis), that of a Gaussian source (RM gradient along the minor axis), or a combination of the two.
If a source consists of uniform and Gaussian components that from our perspective all lie behind the same linear gradient in RM then the frequency and RM spectra are a combination of the spectra which we modelled in \S~\ref{burnjet.sec} and this section.
At high frequencies the uniform source depolarizes at roughly the same rate as a Gaussian source of the same extent, i.e. the long axis of the uniform source has a half-length equal to approximately three times the standard deviation of the Gaussian source.
The grey lines in the background of Fig.~\ref{gauss_jet_spectra.fig} show the spectra of uniform sources with the same extent as the Gaussian sources that we modelled; 
the lines for each matching pair of spectra intersect at about 300 flux density units.
At high frequencies the difference between the spectra of the two source types becomes more pronounced when the RM gradient across the Gaussian and uniform source becomes steeper, while at low frequencies the re-brightening of spectra is a tell-tale sign for uniform sources, which is not found in Gaussian sources.
If the uniform source is larger than the matching Gaussian source it will be more depolarized than the Gaussian source, and vice versa. 

\subsubsection{Cylindrical source with an azimuthal magnetic field}\label{cylinder.sec}
For this geometry, Faraday rotation occurs in a boundary layer between an inner, emitting cylinder of radius $R$ and an outer, coaxial, cylinder of radius $R'$. 
The cylinders can have any inclination with respect to the plane of the sky.
In Appendix~\ref{Appendix_A} we explain in detail how we model frequency spectra for this source type; there we also show that the frequency spectra that we model are independent of the inclination angle.
This implies that the inclination angle of the cylinder can not be derived from the frequency or RM spectrum of a source.
For this simple geometry the only input parameters are the maximum RM of the source, RM$_\mathrm{max}$, which is found at the edges of the emitting cylinder, and the intrinsic polarized flux density that is emitted by the source, which we normalize to 1000 units of polarized flux density.
Our model of the cylinder also acts as a test case for the more complex model of an emitting ellipsoid that we will discuss in \S~\ref{ellipsoid.sec}.
We found that a few of the 9800 frequency channels that we simulated do not decrease monotonically in polarized flux density with decreasing frequency; this could be the result of our numerical integration scheme. Because the difference in polarized flux density is at most 0.2 units of polarized flux density, and often at least one order of magnitude smaller, and because only very few frequency channels are affected, the RM spectra that we calculate from our simulation are accurate down to at least 0.01 units of polarized flux density.

\begin{figure}
\resizebox{\hsize}{!}{\includegraphics{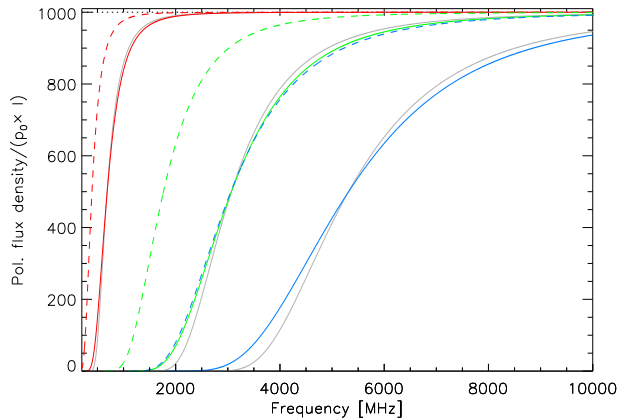}}
\caption{
Frequency spectra for a cylinder of polarized emission with an azimuthal magnetic field which wraps around the major axis of the cylinder. The solid lines indicate cylinders with RM$_\mathrm{max}$ of 25, 500, and 1500 rad~m$^{-2}$ (red to blue) and a Faraday-rotating layer with a thickness equal to 10\% the radius of the emitting cylinder; the dashed lines show results for the same RM$_\mathrm{max}$ but a Faraday-rotating layer with only 1\% the radius of the emitting cylinder. The grey spectra  in the background are for Gaussian sources with a transverse linear RM gradient of 7, 145, and 436 rad~m$^{-2}$~FWHM$^{-1}$ across its major axis. 
}
\label{cyl_spectra.fig}
\label{cyl_gauss_spectra.fig}
\end{figure}

Fig.~\ref{cyl_spectra.fig} shows polarized flux density spectra for different values of RM$_\mathrm{max}$, if Faraday rotation occurs in a boundary layer with a thickness equal to 1\% or 10\% of the radius of the emitting cylinder.
Increasing the thickness of the boundary layer leads to more depolarization, which has the effect of spreading out the curves that correspond to different RM$_\mathrm{max}$ in Fig.~\ref{cyl_spectra.fig}. This makes it easier to tell the curves for the different RM$_\mathrm{max}$ apart, and it makes it easier to determine RM$_\mathrm{max}$ more accurately from observations that cover a wide frequency range. 
Fig.~\ref{cyl_spectra.fig} also shows that spectra of a cylinder with a thick Faraday-rotating layer and a small RM$_\mathrm{max}$ can be very similar to spectra of a cylinder with a thin boundary layer and a large RM$_\mathrm{max}$, which is illustrated by the blue dashed curve which overlaps with the green solid curve.
This will make it difficult to tell the thickness of the boundary layer from the observed frequency spectrum.

In Fig.~\ref{cyl_gauss_spectra.fig} we also compare the frequency spectra for a cylinder with a thick Faraday-rotating layer with frequency spectra for a Gaussian source with a transverse linear RM gradient across its major axis, which we modelled in \S~\ref{gaussjet.sec}. 
We chose the values for the RM gradients in front of the Gaussian source such that frequency spectra of the cylinder and the Gaussian source intersect at 500 polarized flux density units.
The comparison between the coloured and greyscale frequency spectra shows that it could be possible to tell the two source types apart by accurately measuring the difference in curvature of the frequency spectra, which becomes easier if the amount of Faraday rotation across the Gaussian source or across the minor axis of the cylinder is large.

\subsubsection{Ellipsoidal source with an azimuthal magnetic field}\label{ellipsoid.sec}
\begin{figure}
\resizebox{\hsize}{!}{\includegraphics{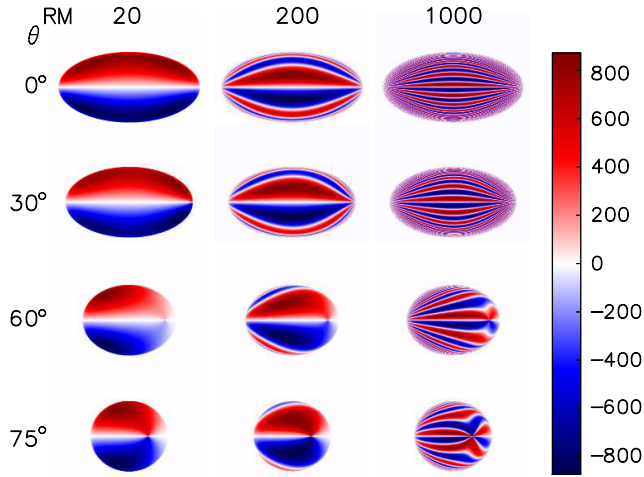}}
\caption{Stokes $Q$ profiles (in arbitrary flux density units) at 1.4 GHz when the surface of a spheroid with axis ratio 1:1:2 and a thick Faraday screen is projected on the sky.
Different rows indicate different inclination angles $\theta$ of the major axis of the spheroid, while different colums show configurations with RM$_\mathrm{max}$ of 20, 200, and 1000 rad~m$^{-2}$, respectively.
}
\label{sphere_configurations.fig}
\end{figure}

In this model, Faraday rotation occurs between an inner and outer ellipsoid, where the magnetic field in the Faraday-rotating layer wraps around the major axis ($z$ axis) of the inner ellipsoid. These ellipsoids have a common coordinate system that we introduced at the beginning of this Section, but their extents can be chosen independently. 
In Appendix~\ref{Appendix_C} we describe in detail how we modelled this source type.
The ellipsoid is described by many parameters; we reduce the dimensionality of the parameter space by considering the case of a spheroid, which has a circular cross-section of radius $R$ perpendicular to the major axis $z$.
The spheroid can be described by specifying the axis ratio $C/R$ between the length of polar ($z$) axis $C$ and its radius at the equator $R$, the inclination $\theta$ of the spheroid with respect to the plane of the sky, the thickness of the Faraday-rotating layer, and RM$_\mathrm{max}$. 
Fig.~\ref{sphere_configurations.fig} shows the pattern of Stokes $Q$ across the surface of a spheroid which has different inclinations with respect to the plane of the sky and different RM$_\mathrm{max}$.

\begin{figure*}
\resizebox{\hsize}{!}{\includegraphics{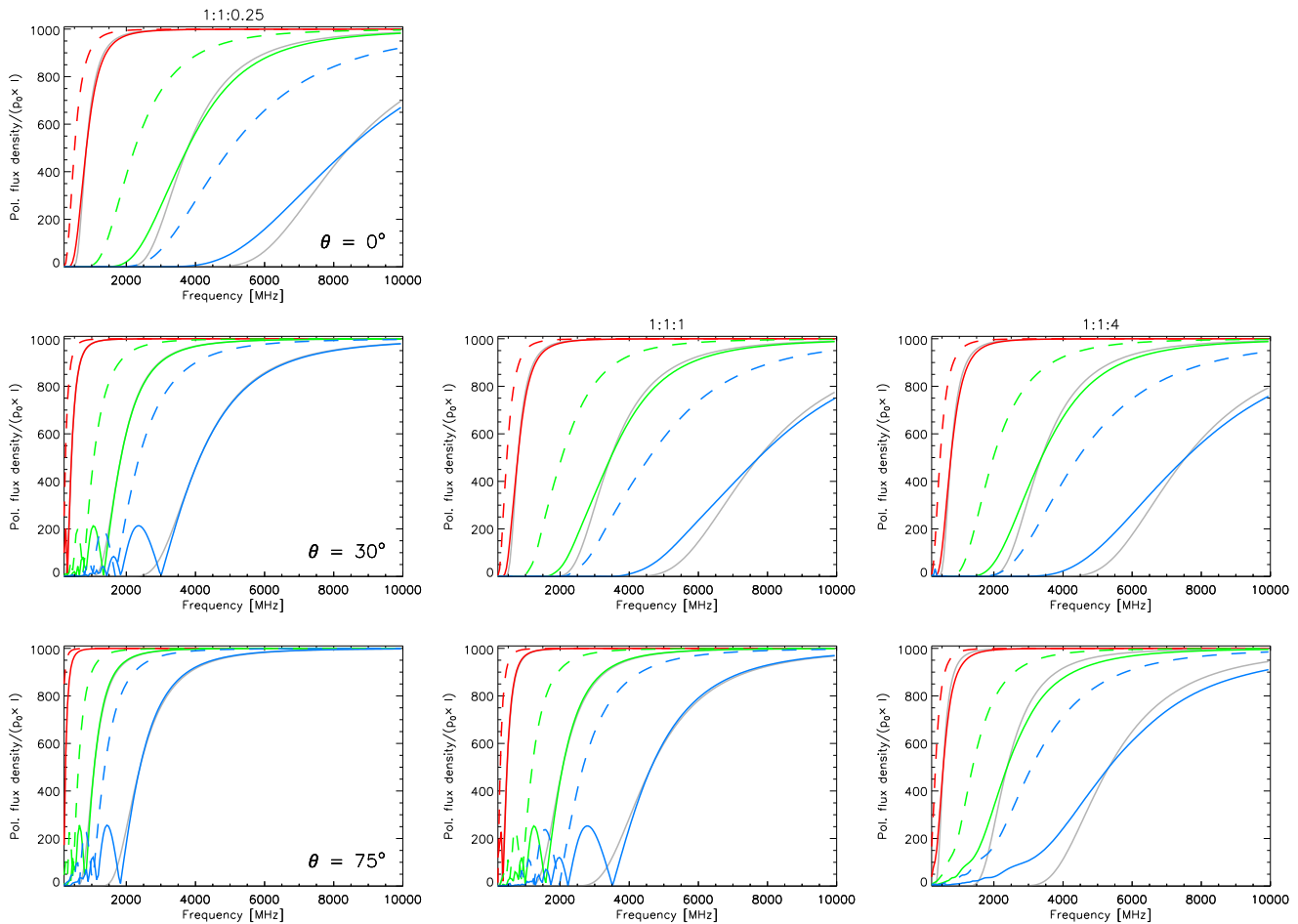}}
\caption{Frequency spectra for different axis ratios of the spheroid (columns), inclination angles $\theta$ (rows), and RM$_\mathrm{max}$ of 25, 500, and 2500 rad~m$^{-2}$ (red to blue).
One in every 25 channels is shown.
The axes of the outer spheroid are scaled versions of the axes of the inner spheroid, where the scale factor is either 110 per cent, producing a thick layer of Faraday-rotating material (solid lines) or only  101 per cent, producing a thin layer (dashed lines).
The axis ratios of each column are indicated above the top panel in each column.
The three panels for a spheroid which lies in the plane of the sky ($\theta$= 0$^\circ$) are almost identical; we show only one of the panels (see the text for details).
The grey lines in the background show frequency spectra for a Gaussian source with a transverse linear RM gradient, which we modelled in \S~\ref{gaussjet.sec}. The RM gradients in front of these Gaussian sources were chosen such that the spectrum of a spheroid and the spectrum of the matching Gaussian source intersect at 500 flux density units.
}
\label{spheroid_thick.fig}
\end{figure*}

Fig.~\ref{spheroid_thick.fig} shows frequency spectra for spheroids with different axis ratios, inclination angles, RM$_\mathrm{max}$, and thicknesses of the layer of Faraday-rotating layer.
The numerical accuracy of these spectra is $\lesssim$ 10 units of polarized flux density; in Appendix~\ref{Appendix_C} we explain how we derived this value.
If the source lies in the plane of the sky the different axis ratios that we tested produce the same frequency spectrum to within a few units of polarized flux density at 500 MHz; therefore we only show one of the three panels for this inclination angle.
Similar to what we noticed for the cylindrical source, increasing the thickness of the Faraday-rotating layer spreads out the frequency spectra.
Changing the inclination of the spheroid clearly has a larger impact on the spectra shown in Fig.~\ref{spheroid_thick.fig} than changing the axis ratio of the spheroid.
Therefore it is easier to determine the inclination of the spheroid than its axis ratio.
If the source is inclined with respect to the plane of the sky frequency spectra can show re-brightening at low frequencies, with secondary maxima which reach the same height as the secondary maxima of the uniform source discussed in \S~\ref{burnjet.sec}.
As we concluded in \S~\ref{burnjet.sec}, this occurs when the emitted polarized flux density is the same for all RM at which the source emits.
Needle-like sources with large $C/R$ axis ratios do not show a strong amplitude difference between the secondary and higher-order maxima and minima.

At high frequencies the spectrum of a spheroid can be approximated by the spectrum of a Gaussian source which we modelled in \S~\ref{gaussjet.sec}.
Accurate measurements at low and intermediate frequencies can detect the secondary and higher-order maxima which are present in the spectrum of an inclined spheroid but not in the spectrum of a Gaussian source; such measurements might even distinguish between an inclined spheroid and a uniform source with a transverse linear RM gradient (\S~\ref{burnjet.sec}).

The spheroids that we modelled do not show a change in position angle with frequency beyond the numerical accuracy of the models.
Therefore, the net RM of the spheroids is zero, which we also found for the models with large-scale magnetic fields which we considered previously.

\subsection{A turbulent foreground medium}\label{turbulent.sec}
Finally, we consider depolarization due to a turbulent magnetic field in front of a uniformly emitting source. We investigate how the coherence length and magnetic field strength affect the polarized flux density spectrum as a function of frequency.

Depolarization by a turbulent foreground has been investigated by a number of authors. \citet{burn1966} considered a foreground screen with RMs that are drawn from a Gaussian distribution with variance $\sigma^2_\mathrm{RM}$. All cells in this model have the same volume equal to the cube of the coherence length of the magnetic field. 
\citet{tribble1991} expanded this analysis to include structure in RM on a spectrum of scales. 
\citet{murgia2004} modelled the polarization properties of galaxy clusters in 3D using a power-law to describe structure in the magnetic field, and showed that the RM dispersion as a function of impact parameter, $\sigma_\mathrm{RM}\left(R_\perp \right)$,  can be approximated well by a model with a single characteristic length scale of the magnetic field\footnote{This length scale is equal to $\int_0^\infty w_\|(R)\mathrm{d}R / w_\|(0)$, where $w_\|(R)$ is the spherically-averaged autocorrelation function of the magnetic field component along the line of sight.}.
\citet{fanti2004} investigated how in Compact Steep Spectrum sources the RM dispersion follows the King profile $\sigma_\mathrm{RM}\left(R_\perp \right) \propto \left(1+ R_\perp^2/R_\mathrm{c}^2\right)^{\left(1-6\beta\right)/4}$ (\citealt{king1962}; also known as a $\beta$ profile) where $R_\mathrm{c}$ is the core radius of the galaxy under investigation.
\citet{rossetti2008} proposed a model where the turbulent foreground only covers a fraction of the emitter in the background. In their model, the observer sees a combination of the Burn depolarization model and a contribution by the background source that is not covered by the turbulent foreground. 
The RMs in their model are drawn from a Gaussian distribution, and there are at most two Faraday-rotating screens with turbulent layers.  
\citet{hovatta2012} modelled depolarization by a turbulent Faraday screen with a Gaussian probability density function (pdf) of RM in front of a radio jet. They concluded that the depolarization of isolated jet components in their sample of AGN can be explained by a small number of sightlines passing through such a Faraday screen.

We model a circular source with a diameter of 25 pc that emits 1000 units of polarized flux density uniformly across its surface.
The results we derive apply also to sources which are not circular, as long as the surface area of those sources is the same as the surface area of the circular source we consider.
The coherence length of the magnetic field is allowed to vary continuously; therefore the number of turbulent cells does not have to be an integer.
The intrinsic polarization angle of the emission is set to zero degrees throughout the emitter. 
We model a Faraday-rotating screen in front of this source that consists of one layer of turbulent cells (\S~\ref{monolayer.sec}) or multiple layers of cells (\S~\ref{multilayer.sec}).
The electron density $n_\mathrm{e}$ is constant throughout the screen, and equal to 10 cm$^{-3}$. 
We investigate magnetic field strengths $B_\mathrm{turb}$ of 1, 5, 10, 25, and 50 $\mu$G, while the direction of the magnetic field is drawn independently between turbulent cells. 
One can choose a different value for the electron density or magnetic field strength; as long as the product of the electron density times the magnetic field strength is the same as in our simulation one will find the same frequency spectra and RM spectra.

\begin{figure*}
\resizebox{0.49\hsize}{!}{\includegraphics{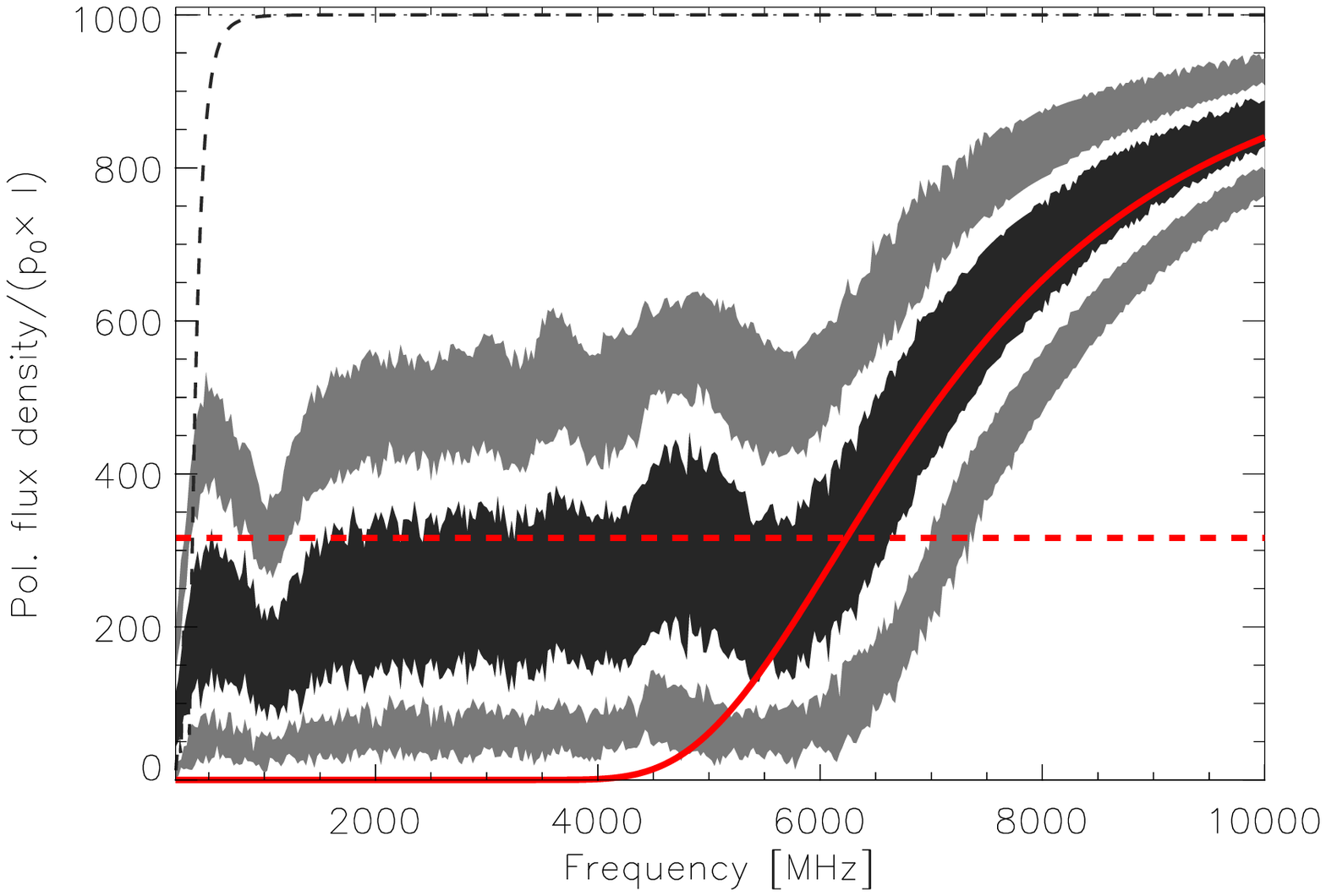}}
\resizebox{0.49\hsize}{!}{\includegraphics{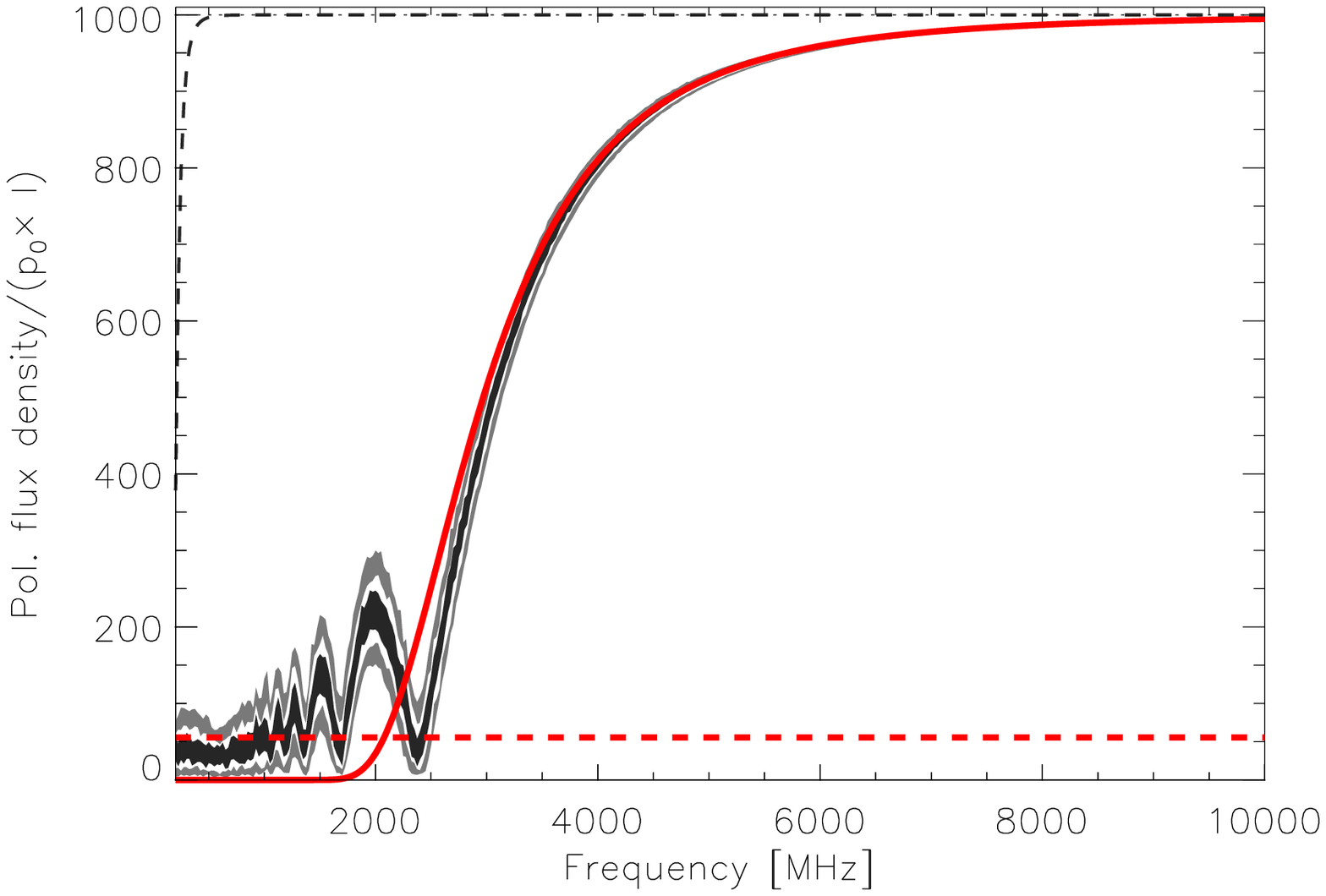}}
\caption{The equivalent of Fig.~\ref{burn_jet_spectra.fig} for a single layer of 10 turbulent cells (left panel) with RMs between $\pm$ 569 rad~m$^{-2}$, and a single layer with 324 turbulent cells (right panel), with RMs between $\pm$ 100 rad~m$^{-2}$. The greytones indicate the different confidence intervals, which contain 95\% (dark grey), 75\% (white) and 50\% (black) of the simulated ensemble. 
These RMs are drawn from a uniform probability density function for RM, and we repeated this process 2000 times to calculate the confidence intervals.
The amplitude of the turbulent magnetic field is 10 $\mu$G, the free electron density 10 cm$^{-3}$, and only one in 25 of the 1-MHz channels is shown. 
The grey dashed line indicates how a single frequency channel is depolarized by the largest possible RM in the Faraday screen, RM$_\mathrm{max}$.
The red line shows the prediction for the length of the monochromatic polarization vector by the depolarization model by \citet{burn1966} who assumed a Gaussian pdf(RM). We calculated the standard deviation of the Gaussian analytically from the pdf(RM) of the monolayer. The red dashed line indicates the root-mean-square monochromatic polarized flux density level that is expected for a two-dimensional random walk model.
}
\label{turbulent_monolayer_spectra.fig}
\end{figure*}

\begin{figure*}
\resizebox{0.49\hsize}{!}{\includegraphics{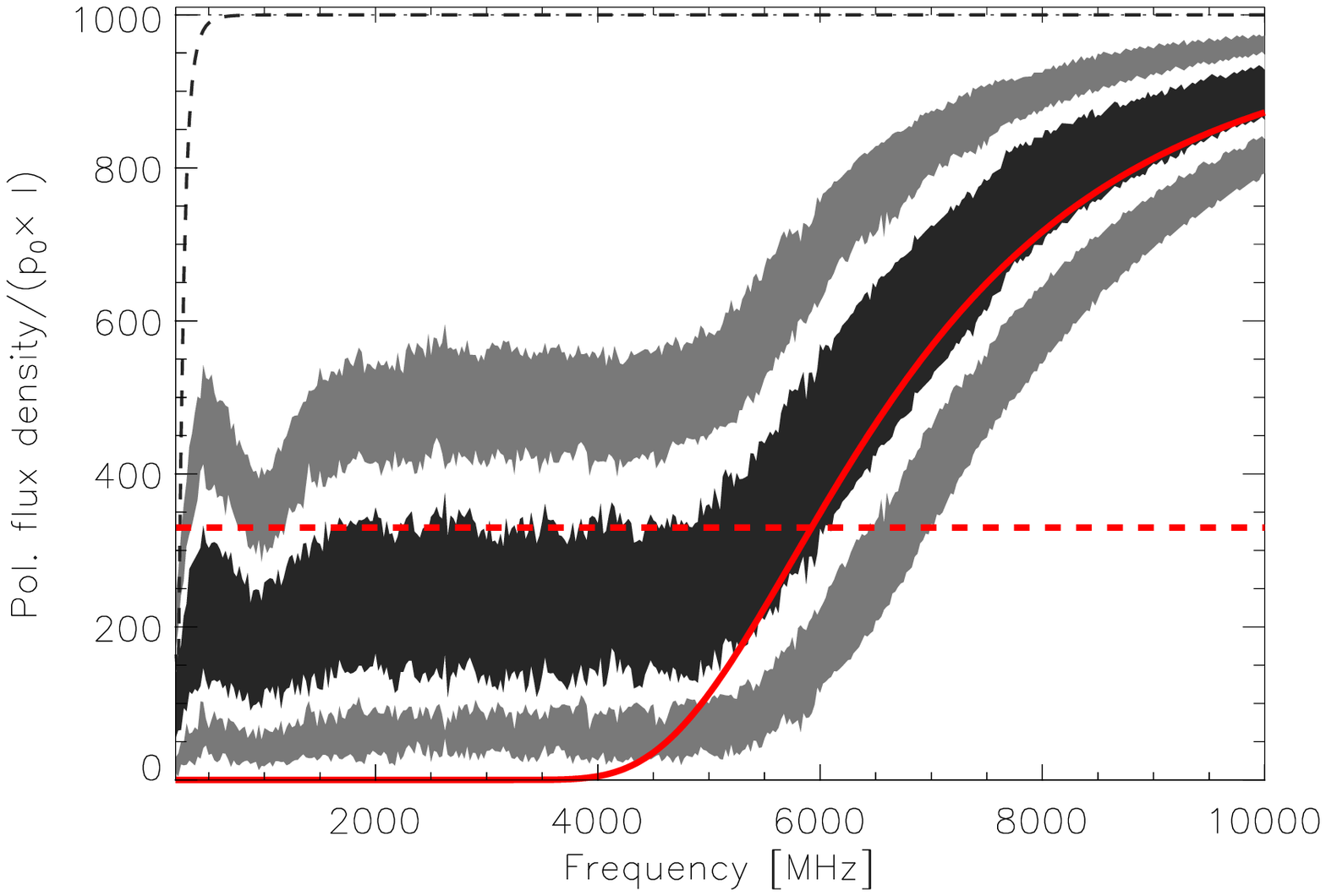}}
\resizebox{0.49\hsize}{!}{\includegraphics{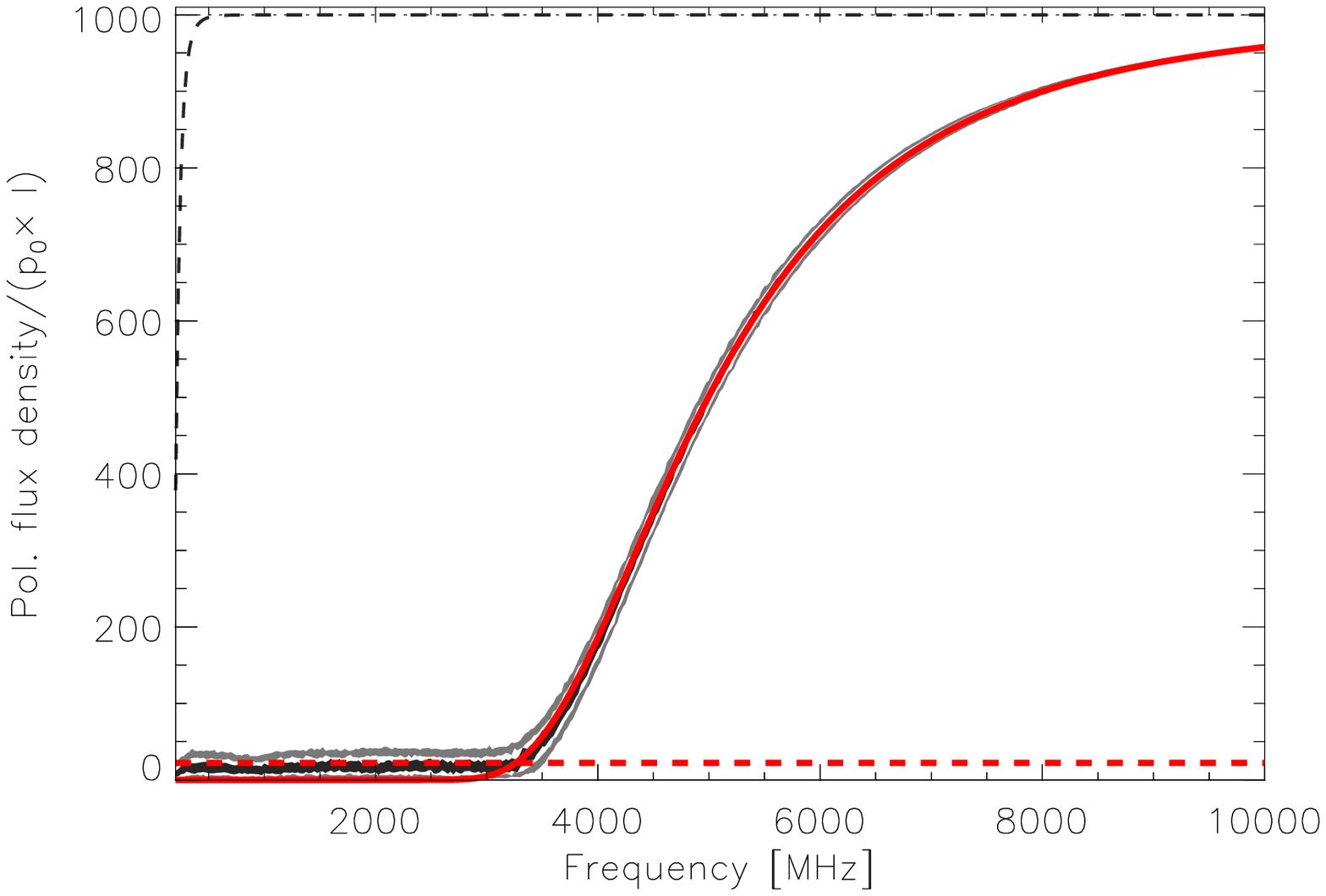}}
\caption{
The equivalent of Fig.~\ref{turbulent_monolayer_spectra.fig} for a turbulent Faraday screen which consists of eight layers of cells. In the panel on the left $B_\mathrm{turb}\ =\ 25\ \mu\mathrm{G}$ divided by 8 (the number of layers), which produces RMs between $\pm$ 1422 rad~m$^{-2}$.
In the panel on the right each layer in the Faraday screen contains 2022 cells; in this case each cell has a magnetic field strength of 25 $\mu$G. 
}
\label{turbulent_eightlayer_spectra_bturb25.fig}
\end{figure*}

\subsubsection{A single layer of cells}\label{monolayer.sec}
For a single layer of turbulent cells the observed monochromatic polarization vector is equal to
\begin{eqnarray}
\bmath{P}_\mathrm{obs}\left(\nu\right)\ =\ \sum_{j=1}^{N_\mathrm{l.o.s.}}\left(\int_{j} \bmath{P}_\mathrm{em}\left(y,z\right)\mathrm{d}y\mathrm{d}z \right)\mathrm{e}^{\mathrm{2iRM_j}\left(\mathrm{c}/\nu\right)^2}\, .
\label{p_obs_turbulent}
\end{eqnarray}
`l.o.s.' is shorthand for `line of sight', there are $N_\mathrm{l.o.s.}$ sightlines, and the coordinate axes $y$ and $z$ point along the minor respectively the major axis of the emission region.
Because the source that we model emits uniformly across its surface, the integral in equation~\ref{p_obs_turbulent} can be replaced by $P/\left(p_\mathrm{0}\times\ I\right) = 1000$ flux density units divided by the number of lines of sight through the Faraday screen. 
This simplifies equation~\ref{p_obs_turbulent} to 
\begin{eqnarray}
\bmath{P}_\mathrm{obs}\left(\nu\right)\ =\ 
 \frac{P/\left(p_\mathrm{0}\times\ I\right)}{N_\mathrm{l.o.s.}} \sum_{j=1}^{N_\mathrm{l.o.s.}} \mathrm{e}^{\mathrm{2iRM_j}\left(\mathrm{c}/\nu\right)^2}\, .
\label{p_monochromatic_turbulent}
\end{eqnarray}
In Appendix~\ref{Appendix_D} we show that if the magnetic field direction is drawn randomly, such that every point on the unit sphere has equal probability of being drawn, then pdf(RM) is uniform between $\pm$ the largest possible RM of the screen,  $\mathrm{RM}_\mathrm{max}$.
In our simulation the maximum RM of a single turbulent cell is given by
\begin{eqnarray}
\mathrm{RM}_\mathrm{max}\ =\  0.81 B_\mathrm{turb} n_\mathrm{e}\sqrt{\upi \left(25/2\,\mathrm{pc}\right)^2/N_\mathrm{l.o.s.}}\, . 
\label{rm_max}
\end{eqnarray}
To simulate a Faraday screen with a single layer of turbulent cells we calculate the monochromatic polarized flux density (equation~\ref{p_monochromatic_turbulent}) by drawing one RM per sightline from a pdf(RM) that is uniform between $\pm$ RM$_\mathrm{max}$. 
We repeat this process 2000 times to build up an ensemble of Monte Carlo realisations.
The frequency channels that we simulate are wide enough that position angles can change by more than $\upi$ radians across a single channel.
To accurately calculate the net polarization vector of discrete channels we calculate for each channel how much the position angle changes due to the turbulent cell with the largest (absolute) RM.
We evaluate equation~\ref{p_monochromatic_turbulent} at forty regularly-spaced frequency intervals for each 2$\upi$ revolution in position angle, and we apply the trapezium rule to each of these intervals to calculate the net polarization vector across each channel.

Fig.~\ref{turbulent_monolayer_spectra.fig} shows polarized flux density spectra for a turbulent monolayer with 10 cells across the surface of the emitter (i.e. a field coherence length of 7 parsec) in the panel on the left, and 324 cells (a coherence length of 1 parsec) in the panel on the right. 
The greytones indicate the most compact confidence intervals that contain 50\%, 75\%, and 95\% of the Monte Carlo realisations (coloured with dark grey, white, and light grey, respectively).
As the grey dashed line in each panel indicates,
depolarization across a single channel becomes important only at frequencies below $\sim$ 1 GHz.

The panel on the left in Fig.~\ref{turbulent_monolayer_spectra.fig} shows that if only a few sightlines pass through the turbulent foreground then one will measure a significant polarized flux density even at low frequencies.
The scatter in the measured polarized flux densities is considerable: 50\% of the realisations lie further than $\sim$ 75 flux density units from the median value.
If many sightlines pass through the turbulent foreground then the resulting frequency spectrum looks similar to that of a uniform source with a transverse linear gradient in RM (\S~\ref{burnjet.sec}) if both sources emit over the same range in RM (panel on the right of Fig.~\ref{turbulent_monolayer_spectra.fig}).
This happens because both source types have the same pdf(RM), which is sampled discretely by the turbulent foreground and continuously by the linear RM gradient.
Increasing $\mathrm{RM}_\mathrm{max}$ leads to strong depolarization even at high frequencies. If both the field strength and the number of sightlines through the Faraday screen are increased, in such a way that  $\mathrm{RM}_\mathrm{max}$ is kept fixed at 100 rad~m$^{-2}$, then pdf(RM) is sampled more continuously, and the sinc-like spectrum in the panel on the right of Fig.~\ref{turbulent_monolayer_spectra.fig} becomes smoother. 

The solid red line in both panels of Fig.~\ref{turbulent_monolayer_spectra.fig} indicates the prediction for a Gaussian pdf(RM), as modelled by \citet{burn1966}, for which we calculated the standard deviation analytically from the uniform pdf(RM) that we simulated. 
At high frequencies the resemblance between the spectra produced by the Burn model and by the turbulent monolayer that we simulated is remarkable. However, at low and intermediate frequencies our model predicts a much higher polarized flux density than the Burn model.

To calculate the monochromatic polarized flux density the Burn model assumes a continuously sampled pdf(RM), which requires a very large number of sightlines through the turbulent Faraday screen. 
The polarization vectors from these sightlines align at the highest frequencies that we simulated, and we detect all 1000 units of polarized flux density which the source emits.
At lower frequencies Faraday rotation becomes important, the different polarization vectors become misaligned because of their different RMs, and this leads to depolarization.
Because the Burn model assumes that the number of sightlines through the turbulent screen is very high, depolarization is complete.
However, in our model of the turbulent screen there are far fewer sightlines than the Burn model assumes. 
All turbulent cells produce polarization vectors of the same length because we assumed that the source emits uniformly across its surface, and because depolarization by a large RM across individual frequency channels only becomes important at very low frequencies, as indicated by the dashed lines in Fig.~\ref{turbulent_monolayer_spectra.fig}.
If these polarization vectors would furthermore have random orientations then their behaviour can be described as a two-dimensional random walk.
The root-mean-square (rms) length of the net (summed) polarization vector is equal to the length of the individual polarization vectors, $\left|\bmath{P}_\mathrm{em}\right|/(p_0\times I)/N_\mathrm{l.o.s.}$, times $\sqrt{N_\mathrm{l.o.s.}}$, which describes the rms distance from the origin in a two-dimensional random walk using unit vectors. 
We indicated this rms length of the net monochromatic polarization vector with a red dashed line in Fig.~\ref{turbulent_monolayer_spectra.fig}. 
It is clear that at low and intermediate frequencies a random walk model with a small number of sightlines produces a much larger (rms) polarized flux density than the Burn model.
The red dashed line lies at the high-end of the black confidence interval in all panels. 
This, combined with the fact that the simulated spectrum shows structure while the spectrum for a random walk model does not, indicates that the random walk model should only be used for obtaining a rough estimate of the expected polarized flux density level at low and intermediate frequencies.

\subsubsection{Multiple layers of cells}\label{multilayer.sec}
While the RM distribution of a single layer of turbulent cells can be modelled easily using a Monte Carlo simulation, drawing RMs for each of the cells individually becomes increasingly more computationally expensive if a Faraday screen consists of multiple layers.
Instead, in Appendix~\ref{Appendix_D} we determine pdf(RM) analytically for up to twenty layers of turbulent cells, and we draw $N_\mathrm{l.o.s.}$ RMs from this pdf, one RM for each sightline. 
We draw RMs from the appropriate pdf(RM) using the technique of rejection sampling, which is described in section~7 in \citet{press1992}.
The monochromatic polarization vector $\bmath{P}_\mathrm{obs}\left(\nu\right)$ can then be calculated from equation~\ref{p_monochromatic_turbulent}, using the RMs we draw from pdf(RM) as RM$_j$. 
We use the same method as for the monolayer to integrate over the finite width of the frequency channels.
We assume that different layers are not shifted with respect to each other, so that cells in different layers lie on top of each other. 

Fig.~\ref{turbulent_eightlayer_spectra_bturb25.fig} shows frequency spectra for Faraday screens with eight layers of turbulent cells. 
Spectra for turbulent screens consisting of one, two, four, and eight layers of turbulent cells behave in a similar way: there is a gradual drop-off at the highest frequencies, followed by a flattening at the lowest frequencies.
Spectra of screens that consist of one or two layers of turbulent cells show secondary and higher-order maxima (Fig.~\ref{turbulent_monolayer_spectra.fig}); 
such features are absent from turbulent screen which consist of at least four layers of cells.
At high frequencies the spectrum of a Faraday screen which consists of many layers of turbulent cells closely resembles the spectrum of a screen with a Gaussian pdf(RM), shown as a red solid line in Figs~\ref{turbulent_monolayer_spectra.fig} and \ref{turbulent_eightlayer_spectra_bturb25.fig}. 
This similarity of the spectra is particularly striking if the number of sightlines through the Faraday screen is large, as illustrated by the panel on the right of Fig.~\ref{turbulent_eightlayer_spectra_bturb25.fig}.
Surprisingly, pdf(RM) which are clearly not Gaussian (Appendix~\ref{Appendix_D}) can produce frequency spectra which are similar to the spectrum of a Gaussian pdf(RM).

The shape of pdf(RM) is not important to describe a random walk process; what matters is that the direction of each polarization vector is drawn independently from the same parent distribution as the other vectors.
Therefore we can apply the random walk model that we introduced in \S~\ref{monolayer.sec} also to turbulent Faraday screens which consist of more than one layer of cells.
The red dashed line in Fig.~\ref{turbulent_eightlayer_spectra_bturb25.fig} indicates the rms length of the net monochromatic polarization vector. 
As we found for the monolayer, if the number of sightlines through the Faraday screen is small our random walk model predicts a higher polarized flux density at low and intermediate frequencies than the Burn depolarization model.
The predicted polarized flux density lies at the high-end of the black confidence interval, and should be used as a rough estimate of the actual polarized flux density.

The spectra of many types of astrophysical sources can be described at low and intermediate frequencies by the random walk model we propose.
Models where a background source is only partially covered by a thick layer of Faraday-rotating turbulent cells have been used in the past to explain the depolarization behaviour of compact AGN (e.g., \citealt{rossetti2008}) and of sources with Mg\,{II} absorbers in the foreground (\citealt{bernet2012}) at low radio frequencies. 
The amplitude of the monochromatic polarized flux density in these models is given by
\begin{eqnarray}
\lefteqn{ \left|\bmath{P}_\mathrm{obs}\left(\nu\right)\right|/\left(p_0\times I\right) =  
 f_c\, \mathrm{exp}\left(-2\, \sigma_\mathrm{RM}^2\left(\mathrm{c}/\nu\right)^4\right) + \left(1-f_c\right)
 \, , }
\label{pi_partial_coverage.eqn}
\end{eqnarray}
where $f_c$ indicates the fraction of the polarized flux density which passes through a Faraday screen with a Gaussian pdf(RM); this Gaussian pdf(RM) has a standard deviation $\sigma_\mathrm{RM}$. 
For a uniformly emitting source $f_c$ is equal to the surface area of the background source that is covered by this Faraday-rotating screen.
Also \citet{burn1966} considered the possibility that there are only a few Faraday-rotating `clouds' along the line of sight towards the background source, so that the background source is only partially covered by the Faraday screen.
Partial coverage models predict a slow drop-off when going from high to intermediate frequencies, which levels off towards even lower frequencies; our Monte Carlo model shows a similar behaviour in Figs.~\ref{turbulent_monolayer_spectra.fig} and \ref{turbulent_eightlayer_spectra_bturb25.fig}. 
The polarized flux density of the plateau in the spectrum of a partial coverage model can be predicted also by a 2D random walk model of the polarization vectors.
However, partial coverage models do not predict the sharp drop-off in polarized flux density that can be seen in our models at very low frequencies, which is the result of depolarization across individual frequency channels.
Observations at these frequencies can distinguish between a partial coverage model and the Monte Carlo model we proposed.
In \S~\ref{rm_spectra_turbulent} we show that RM spectra can also be used to tell these models apart. 

\subsection{Comparison between models with large-scale and turbulent magnetic fields}\label{RMdiscussion.sec}
In \S~\ref{channel_depolarization_structured_screen} we showed that the spectra of all the models with large-scale magnetic fields which we considered can be approximated at high frequencies by the spectrum of a Gaussian source with a transverse linear gradient in RM. 
Spectra of sources with turbulent Faraday screens can be approximated at high frequencies by the spectrum of a turbulent screen with a Gaussian pdf(RM) which is pierced by many sightlines (\S~\ref{turbulent.sec}).
In fact, such a turbulent Faraday screen produces the same spectrum as the Gaussian source from \S~\ref{channel_depolarization_structured_screen}.
The Gaussian pdf(RM) of this turbulent Faraday screen is sampled continuously, therefore 
\begin{eqnarray}
\bmath{P}_\mathrm{obs}\left(\nu\right) & = & 
\int_{-\infty}^{\infty} \left(p_0\times I\right)\mathrm{pdf}\left(\mathrm{RM}\right)\mathrm{e}^{2\mathrm{iRM}\left(\mathrm{c}/\nu\right)^2}\mathrm{dRM} \nonumber\\
& = & \frac{p_0\times I}{\sqrt{2\upi}\sigma}\int_{-\infty}^{\infty} \mathrm{e}^{-\frac{1}{2}\left(\mathrm{RM}/\sigma\right)^2}\mathrm{e}^{2\mathrm{iRM}\left(\mathrm{c}/\nu\right)^2}\mathrm{dRM} \nonumber\\
& = & \left(p_0\times I\right)\mathrm{e}^{-2\sigma^2\left(\mathrm{c}/\nu\right)^4}\, , 
\label{gaussian_pdf_spectrum}
\end{eqnarray}
as shown by \citet[][equation~21 in his paper]{burn1966}.  
Equation~\ref{gaussian_pdf_spectrum} depends on frequency in the same way as equation~\ref{rm_gradient_plus_gaussian}, modulo a term in the latter equation which expresses Faraday rotation of the emission.

Determining from observations at high frequencies whether a source has a large-scale or turbulent magnetic field is therefore not straightforward.
One can tell the different models apart by accurately measuring the shape of the spectrum at high frequencies, and by including measurements  at intermediate and low frequencies if available.
One advantage of the similarity of spectra at high frequencies is that the model of a Gaussian source can be used to calculate the approximate shape of the spectrum of a more complex source very quickly.

\begin{table}
\centering
\caption{FWHM of the RM spread function for each of the frequency windows for which we calculate RM spectra. These FWHM values were calculated using equation~61 in \citet{brentjens2005}.
}
\begin{tabular}{r@{ -- }lc}
\hline
\multicolumn{2}{c}{Frequency band} & FWHM\\
\multicolumn{2}{c}{(MHz)} & (rad~m$^{-2}$)\\
\hline
350 & 900 & 6 \\
950 & 1760 & 54 \\
1300 & 3100 & 87 \\
5000 & 7000 & 2158 \\
\hline
\end{tabular}
\label{fwhm.tab}
\end{table}

\section{Rotation measure spectra}\label{sim-rm-spec}
In this Section we simulate RM spectra for four frequency bands: 350--900, 950--1760, 1300--3100, and 5000--7000 MHz, to understand what can be learned about sources that emit over different RM ranges. The first two bands are proposed for SKA1-mid \citep{dewdney2013}, 
the third band is available with the Australia Telescope Compact Array (ATCA), while the fourth band is available with a number of radio telescopes. Each time we will use 1 MHz frequency channels that have uniform response functions.
For each of these frequency windows we provide full-width at half maximum (FWHM) values of the RM spread function in Table~\ref{fwhm.tab}.

Because of the low frequencies and relatively wide frequency channels that we simulate we have to check whether we can use the formalism by \citet{brentjens2005} to calculate RM spectra, or the new formalism that we proposed in an accompanying paper (\citealt{schnitzeler2015}; we will refer to this paper as `SL15'). 
In SL15 we show that the discrete Fourier transform between wavelength squared and RM that is commonly used to calculate RM spectra is only approximately correct for most data sets.
Exact RM spectra can be calculated only if the channel response function of the data is included in this calculation.
The complex exponential exp$\left(-2\mathrm{i}\, \mathrm{RM'}\lambda^2_\mathrm{c}\right)$, where `$\mathrm{RM'}$' is the trial RM and $\lambda^2_\mathrm{c} = \left(\lambda_1^2+\lambda_2^2\right)/2$ is the average wavelength squared of the channel, is exact if frequency channels have a top-hat response function in wavelength squared.
In SL15  we derive how RM spectra can be calculated for frequency channels with any type of channel response function, and we provide expressions for calculating RM spectra if the channel response function has a top-hat shape in frequency.
Many geometries that we simulate emit over a range in RM that is much wider than the FWHM of the RM spread function (RMSF); to add the contributions by emission at (very) different RMs the wings of the RMSF have to be calculated exactly, which requires the formalism that we developed in SL15. 
Therefore we will use in all cases the formalism from SL15, in particular equation~9 from that paper and the normalized version of equation~12, to calculate RM spectra from the frequency spectra that we simulated.

\begin{figure*}
\resizebox{\hsize}{!}{\includegraphics{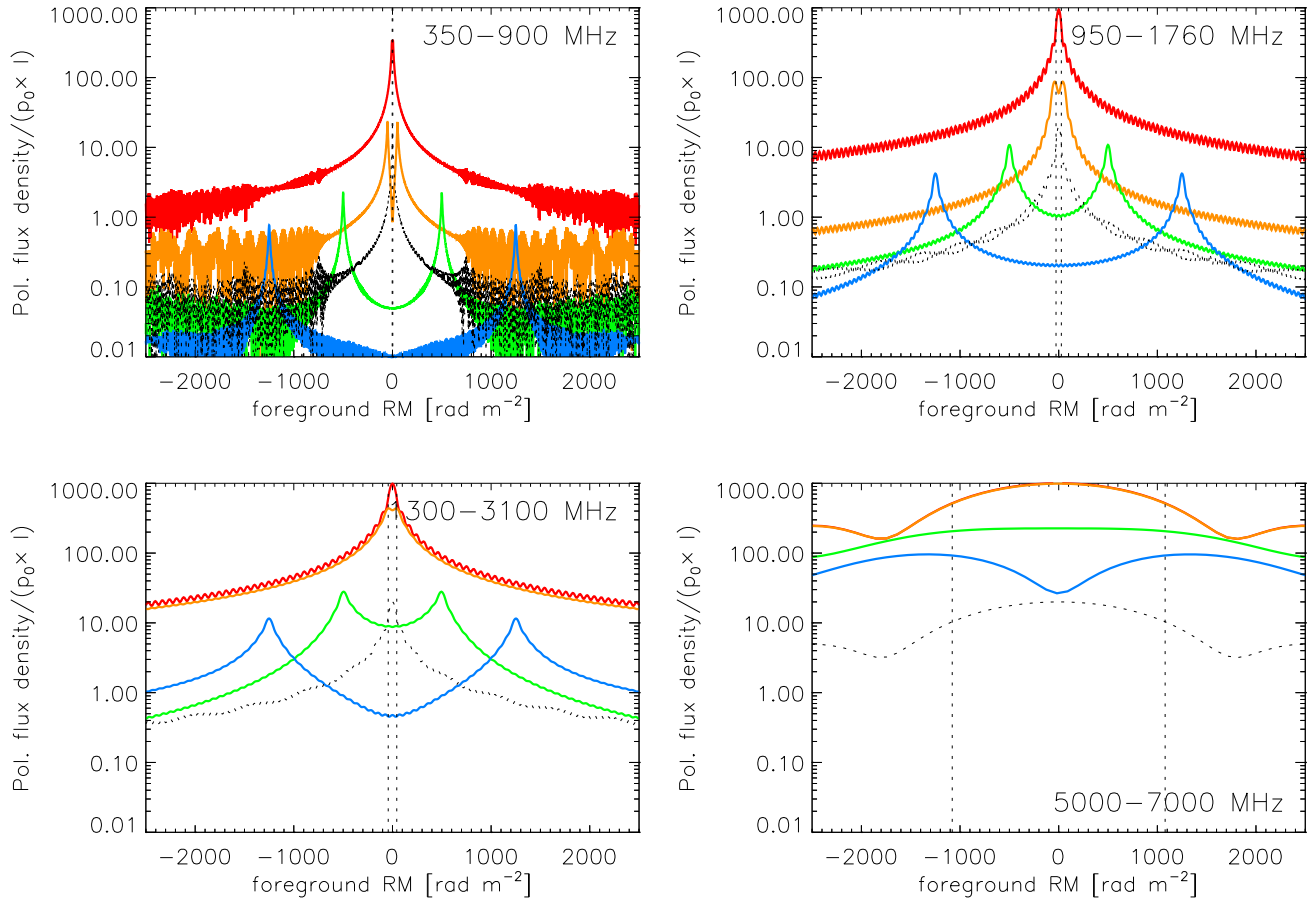}}
\caption{RM spectra for a uniform source with a transverse linear RM gradient from \S~\ref{burnjet.sec}, which emits over an RM range of 10, 100, 1000, and 2500 rad~m$^{-2}$ (red to blue). 
The frequency window used in the computation of RM spectra is indicated in the corner of each panel. The vertical dotted lines indicate the FWHM of the RM spread function as a crude way for estimating when the RM spectrum is resolved. The black spectrum shows the level of instrumental polarization if the source has an intrinsic polarization of 5\% and the polarization purity of the instrument is -30 dB of Stokes $I$. 
The step size in the RM spectra is equal to ten samples per FWHM of the RM spread function for the three lowest frequency windows, and twenty samples per FWHM for the window at the highest observing frequencies.
}
\label{burn_jet_rm_spectra.fig}
\end{figure*}

\begin{figure*}
\resizebox{\hsize}{!}{\includegraphics{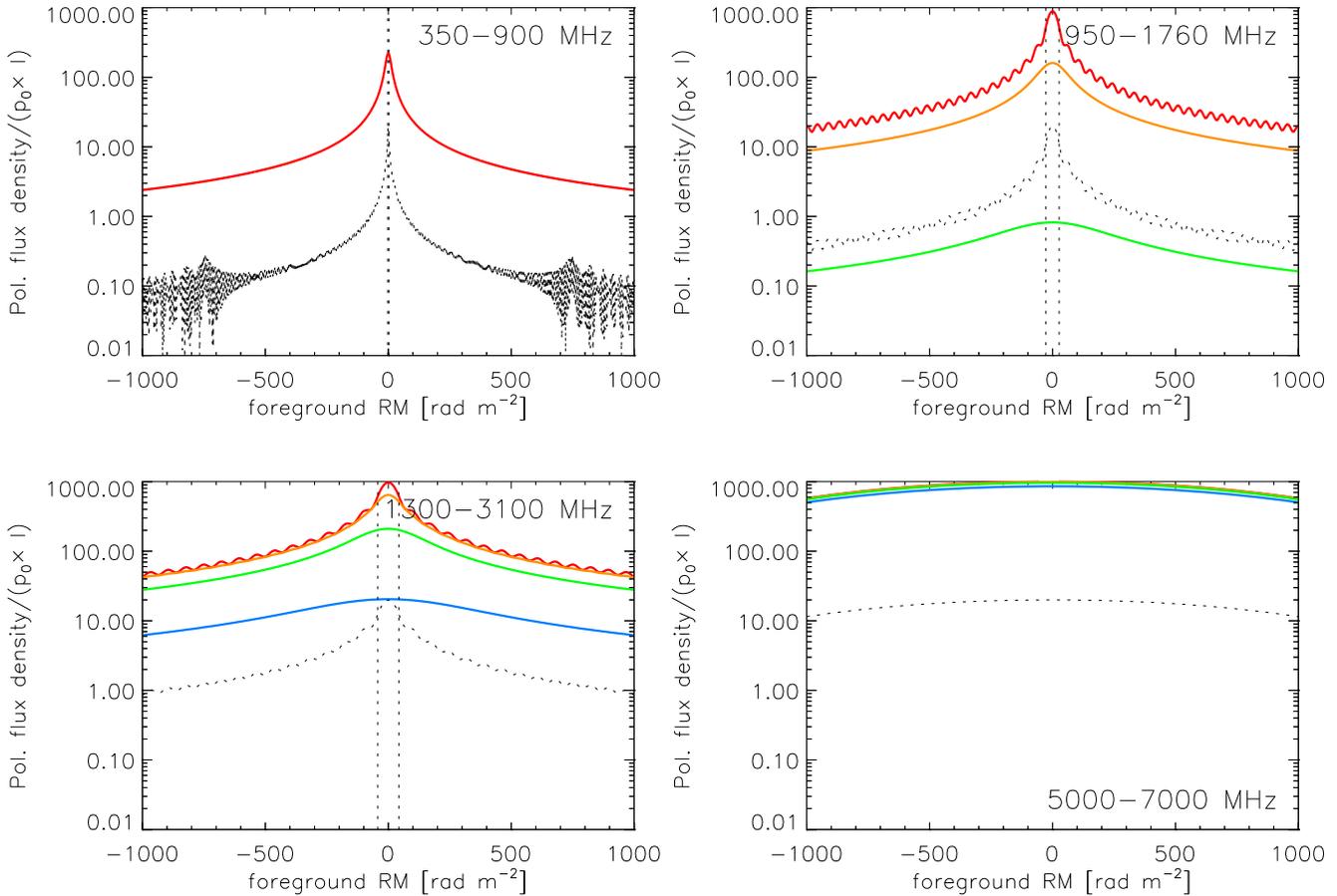}}
\caption{RM spectra for Gaussian sources with RM gradients of 10, 50, 125, and 250 rad~m$^{-2}$~FWHM$^{-1}$ (red to blue), shown in an equivalent way to Fig.~\ref{burn_jet_rm_spectra.fig}. We show a smaller range in RM along the x-axis because the Gaussian sources that we modelled emit over a smaller range in RM than the uniform sources from Fig.~\ref{burn_jet_rm_spectra.fig}. 
}
\label{gauss_jet_rm_spectra.fig}
\end{figure*}

\subsection{Instrumental Polarization}
Sources that are severely depolarized require long integration times to be detected. In addition, polarization leakage from the instrument restricts which polarized sources can be detected even if the observing time is sufficient. 
We model the contribution of instrumental polarization to the RM spectrum as an RMSF centred on RM = 0 rad~m$^{-2}$ with a peak amplitude of -30 dB times the total intensity signal. 
We will assume that the source has an instrinsic polarization percentage $p_0$ = 5\%, which produces an instrumental polarization response with a height of 20 flux density units.
The SKA baseline design specifies a polarization purity of -30 dB across the FWHM of the primary beam \citep{dewdney2013}. This performance is better than the instrumental polarization level of either the Dominion Radio Astrophysical Observatory (DRAO) synthesis telescope and the Very Large Array (VLA) prior to its upgrade to the JVLA (\citealt{taylor2007}, \citealt{condon1998}). 

\subsection{Synthetic RM spectra}
\subsubsection{Geometries with large-scale magnetic fields}\label{rm_spectra_regular}
In Figs~\ref{burn_jet_rm_spectra.fig} and \ref{gauss_jet_rm_spectra.fig} we show RM spectra that we calculated for the uniform source (\S~\ref{burnjet.sec}) and the Gaussian source (\S~\ref{gaussjet.sec}) for four frequency windows. As we discussed in the previous Section, the frequency spectra of the sources that we modelled can be approximated by the spectrum of a Gaussian source, and at low and intermediate frequencies sometimes by the spectrum of a uniform source. 
We show the FWHM of the RM spread function as a rough indication for when the RM spectrum of a source becomes resolved. 

Comparing these two models has the additional advantage that we can investigate how RM synthesis handles the sharp edges of the RM spectrum of the uniform source, while the RM spectrum of the Gaussian source has smooth edges. 
As SL15 showed, for moderate RMs RM synthesis can be approximated well by the discrete Fourier transform that was proposed by \citet{brentjens2005}. 
The rectangular window function then acts as a highpass filter in this Fourier transform, and structure on large RM scales will be missing from the reconstructed RM spectrum of the uniform source, similar to the `missing short spacings' problem in radio interferometry.
Because the Gaussian source that we modelled has a smooth RM spectrum its reconstructed RM spectrum suffers much less from this effect. 
In the lowest frequency window the uniform source and the RM spread function become ragged far from 0 rad~m$^{-2}$, the mean RM of the models. This is the result of the discrete frequency sampling and the frequency channel response functions that we simulated.

Observations in the lowest frequency band provide the most accurate RM measurements, which makes it possible to correctly identify sources that emit over RM ranges that are only slightly different.
Fig.~\ref{burn_jet_rm_spectra.fig} illustrates this for example for the two sources that emit over a range in RM out to $\pm$ 10 rad~m$^{-2}$ and $\pm$ 100 rad~m$^{-2}$, which are difficult to tell apart from observations in the 1300--3100 MHz window, but can be identified correctly from observations in the two windows at lower frequencies.
If sources that emit over a wider range in RM have a lower peak polarized flux density (equation~\ref{sum_P_constant}), then our simulations show that at low frequencies only sources that emit over a small range in RM can be detected above the instrumental polarization level. 
Uniform sources that emit over a wide range in RM might show up only as two peaks above the instrumental polarization threshold.
This effect is less severe at higher frequencies; therefore we recommend including high-frequency data ($\gtrsim$ 1 GHz) to search for sources which emit over a range in RM that is larger than several tens of rad~m$^{-2}$.
The 5--7 GHz band is suitable for identifying sources which emit over RM ranges of more than several hundred rad~m$^{-2}$. 
The key role of this frequency band is to determine the amount of wavelength-independent depolarization, because Faraday rotation is much less severe at high frequencies.

\begin{figure*}
  \begin{minipage}[b]{0.5\linewidth}
    \centering
    \resizebox{\hsize}{!}{\includegraphics[width=\linewidth]{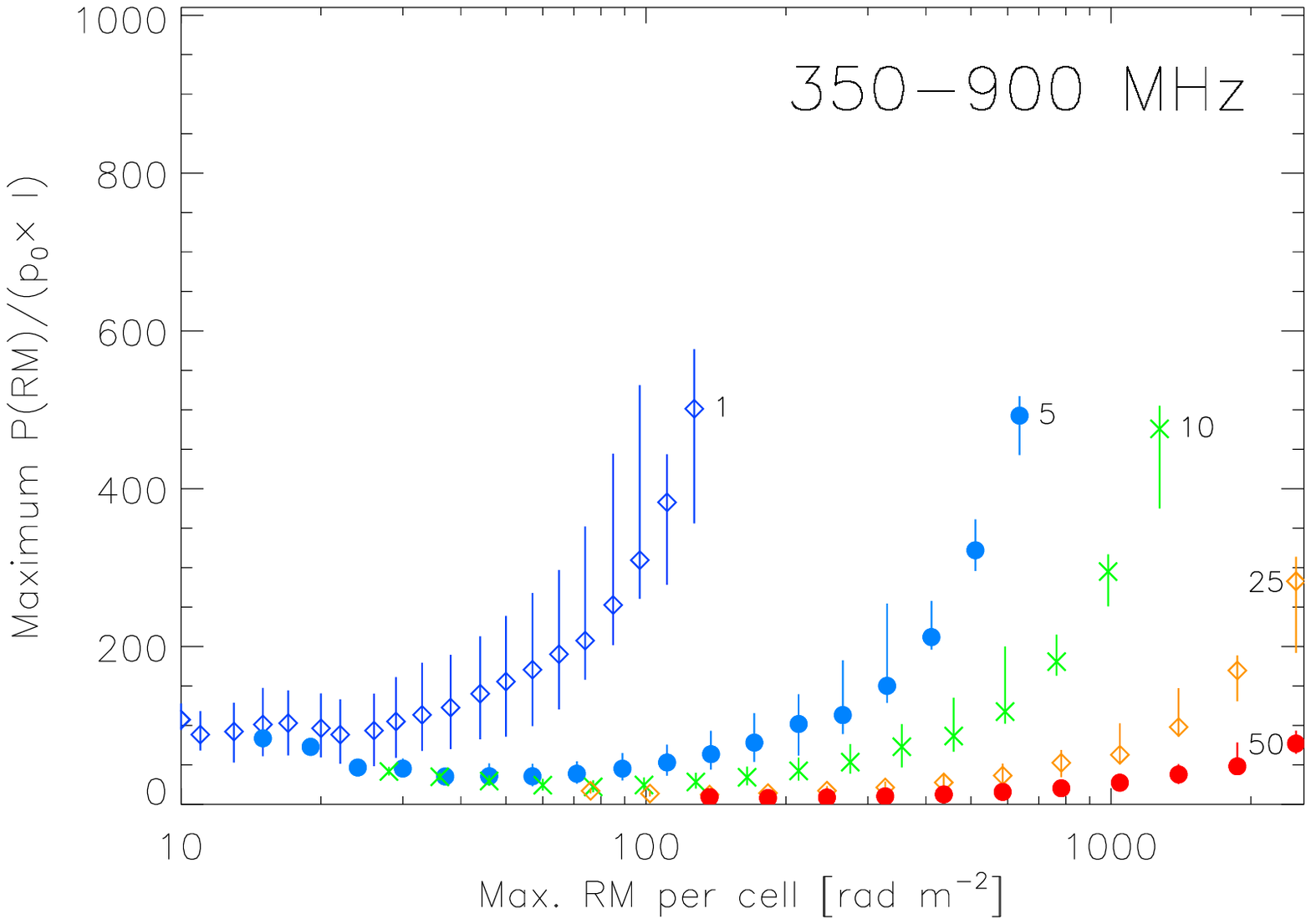}} 
    \vspace{-3ex}
  \end{minipage}
  \begin{minipage}[b]{0.5\linewidth}
    \centering
    \resizebox{\hsize}{!}{\includegraphics[width=\linewidth]{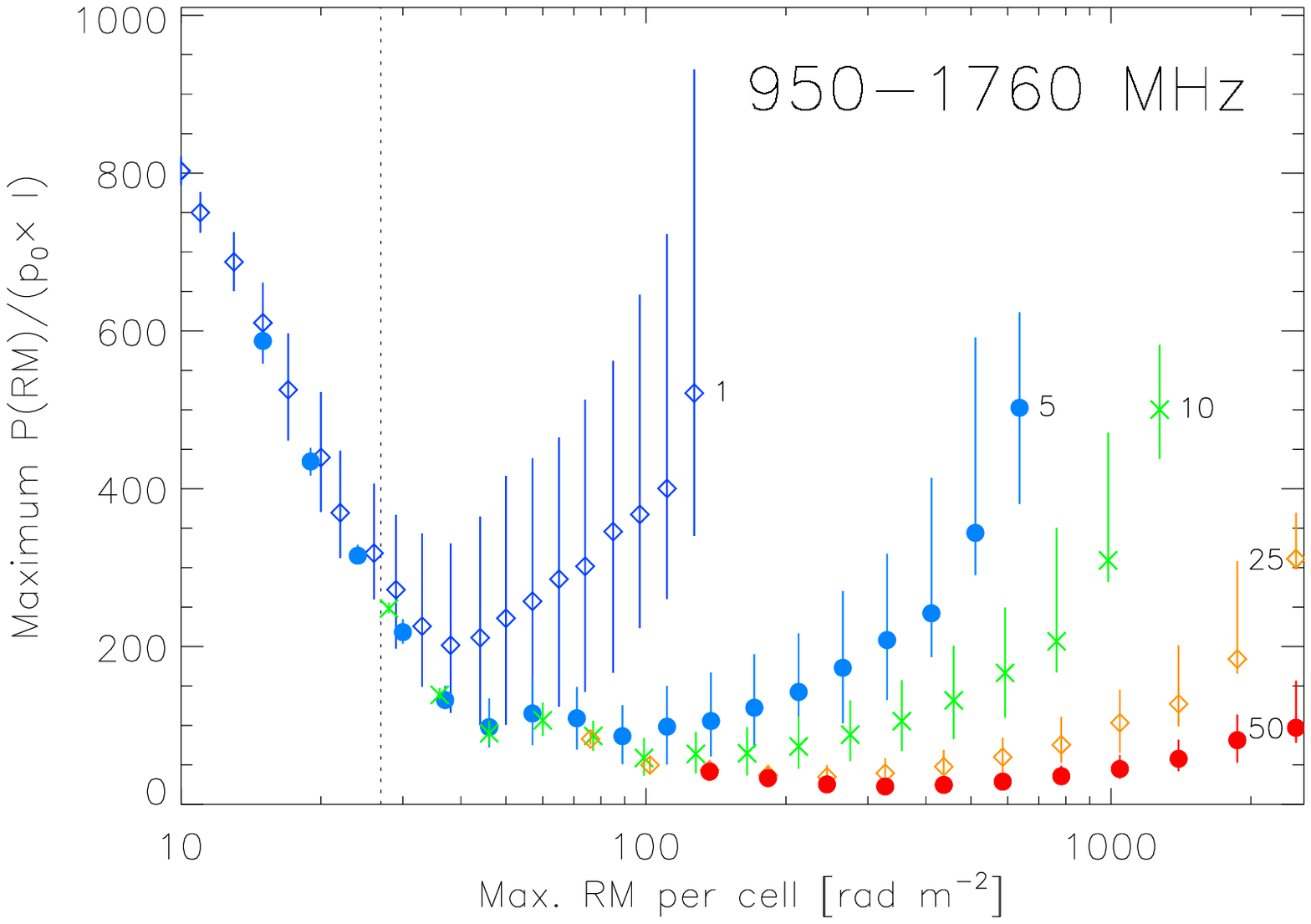}} 
    \vspace{-3ex}
  \end{minipage} 
  \begin{minipage}[b]{0.5\linewidth}
    \centering
    \resizebox{\hsize}{!}{\includegraphics[width=\linewidth]{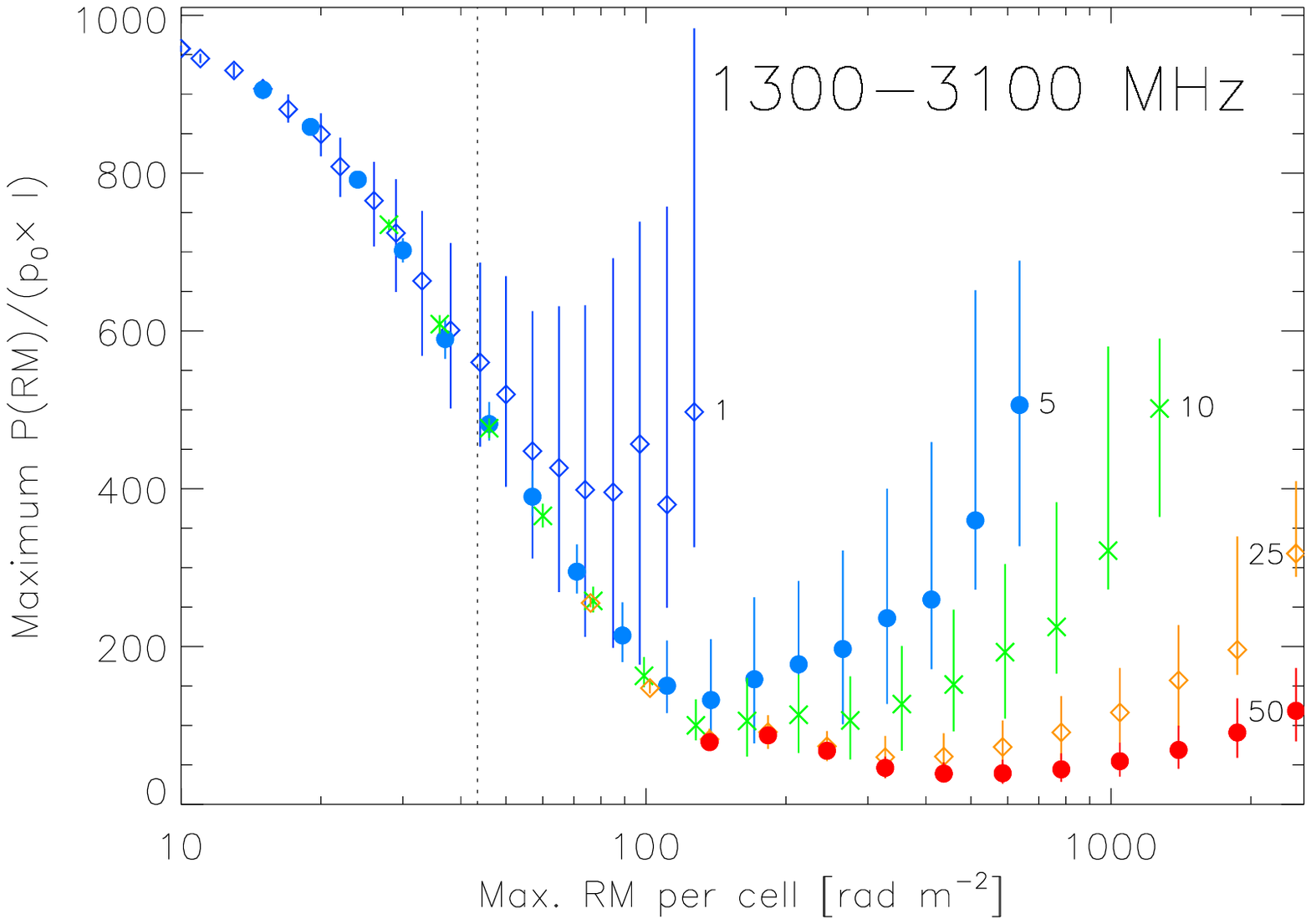}} 
    \vspace{-3ex}
  \end{minipage}
  \begin{minipage}[b]{0.5\linewidth}
    \centering
    \resizebox{\hsize}{!}{\includegraphics[width=\linewidth]{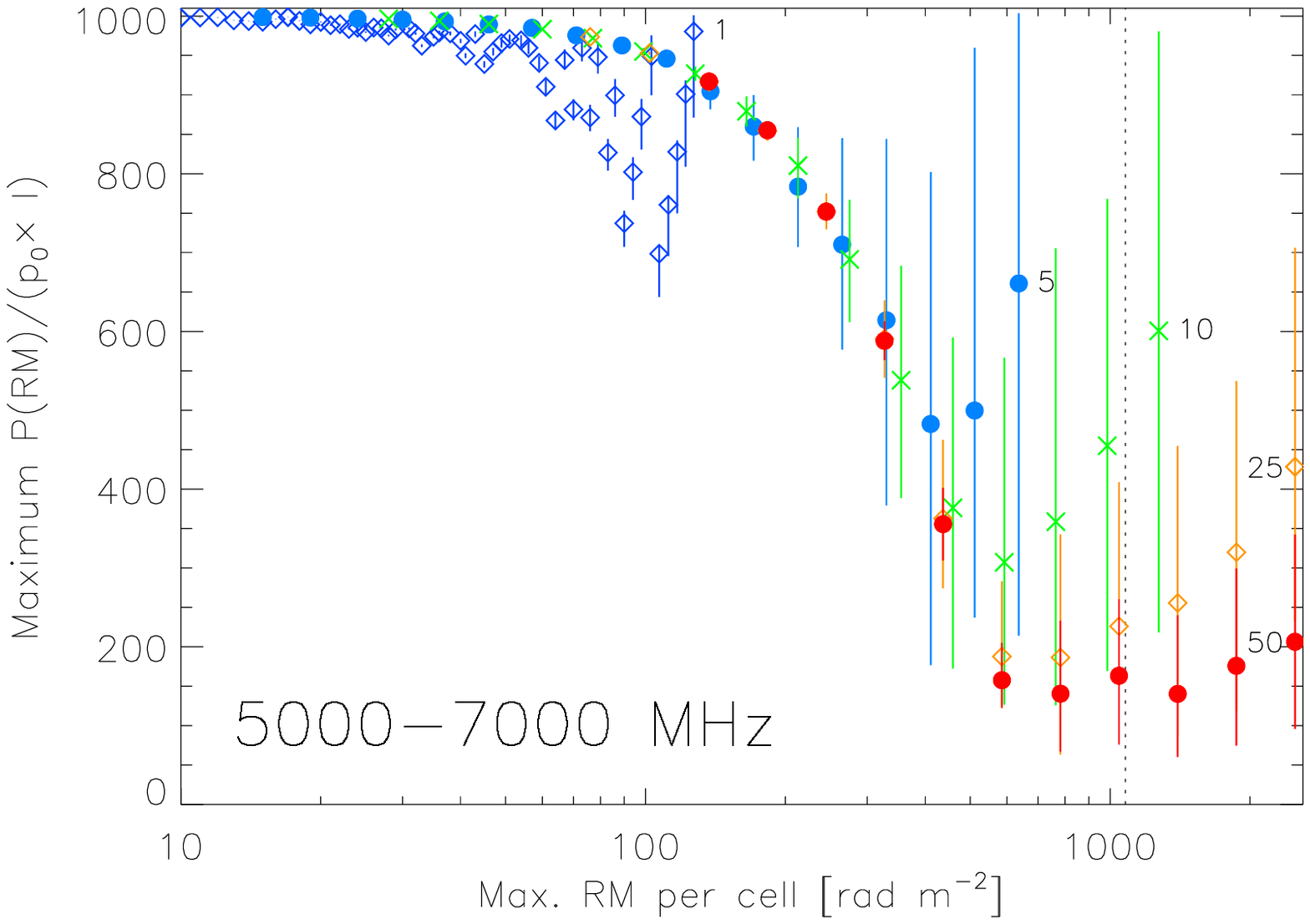}} 
    \vspace{-3ex}
  \end{minipage} 
\caption{Peak polarized flux density in the RM spectrum versus the largest possible RM for a single turbulent cell for the geometry described in \S~\ref{turbulent.sec}, for different frequency windows. In each panel the different colours and plot symbols indicate different values for the magnetic field strength: 1, 5, 10, 25, and 50 $\mu$G. 
The vertical dotted line indicates the half-width at half maximum of the RM spread function from Table~\ref{fwhm.tab}.
When following a curve that connects points with the same colour and plot symbol the coherence length of the magnetic field decreases when going from right to left and the number of sightlines through the Faraday screen increases.
We ran 2000 Monte Carlo simulations to determine the distribution of peak flux densities in the RM spectrum. The plot symbol indicates the median value of this distribution, and the error bars enclose the most compact 95\% confidence interval. 
}
\label{turb_monolayer.fig}
\end{figure*}

\begin{figure*}
  \begin{minipage}[b]{0.5\linewidth}
    \centering
    \resizebox{\hsize}{!}{\includegraphics[width=\linewidth]{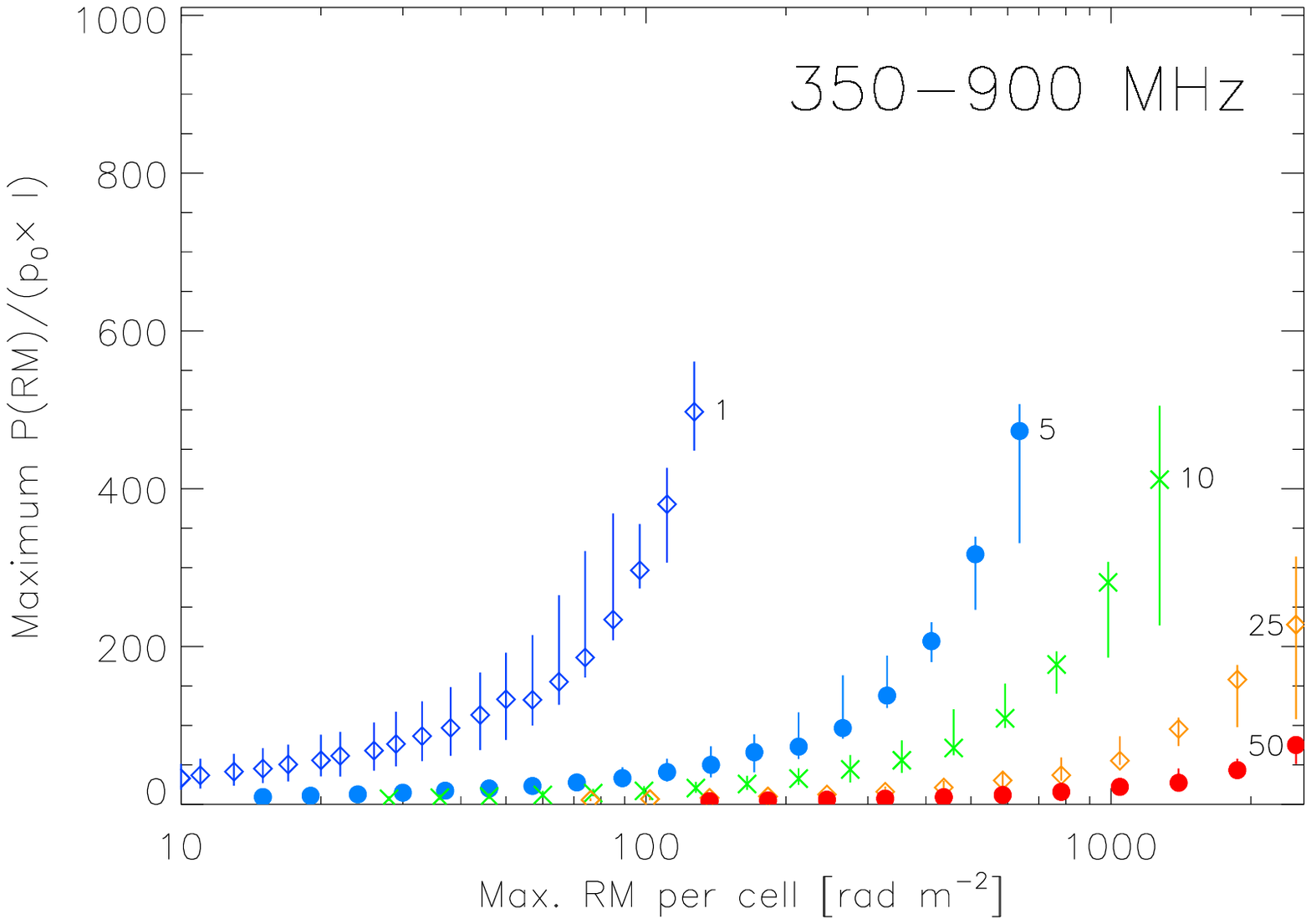}} 
    \vspace{-3ex}
  \end{minipage}
  \begin{minipage}[b]{0.5\linewidth}
    \centering
    \resizebox{\hsize}{!}{\includegraphics[width=\linewidth]{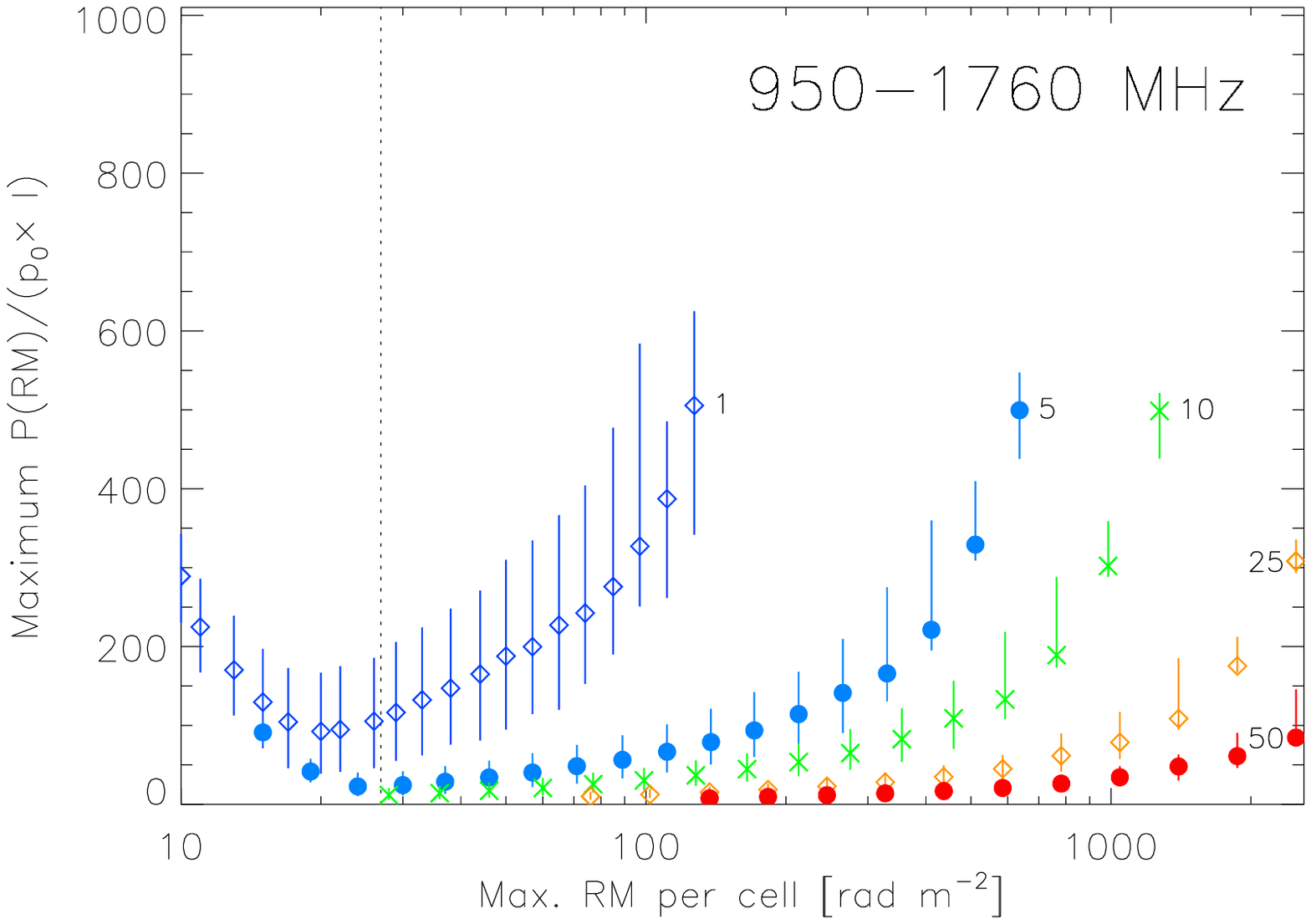}} 
    \vspace{-3ex}
  \end{minipage} 
  \begin{minipage}[b]{0.5\linewidth}
    \centering
    \resizebox{\hsize}{!}{\includegraphics[width=\linewidth]{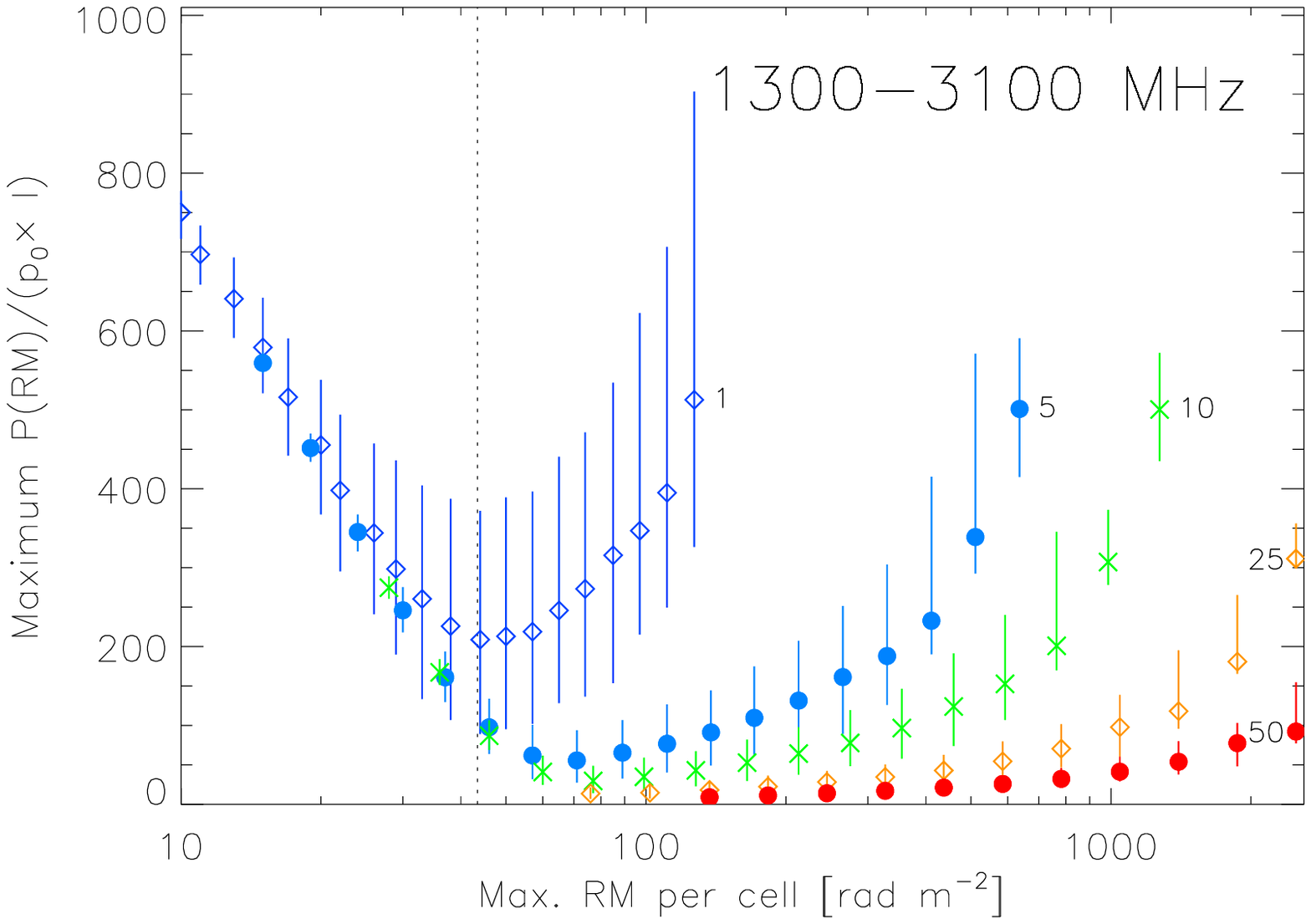}} 
    \vspace{-3ex}
  \end{minipage}
  \begin{minipage}[b]{0.5\linewidth}
    \centering
    \resizebox{\hsize}{!}{\includegraphics[width=\linewidth]{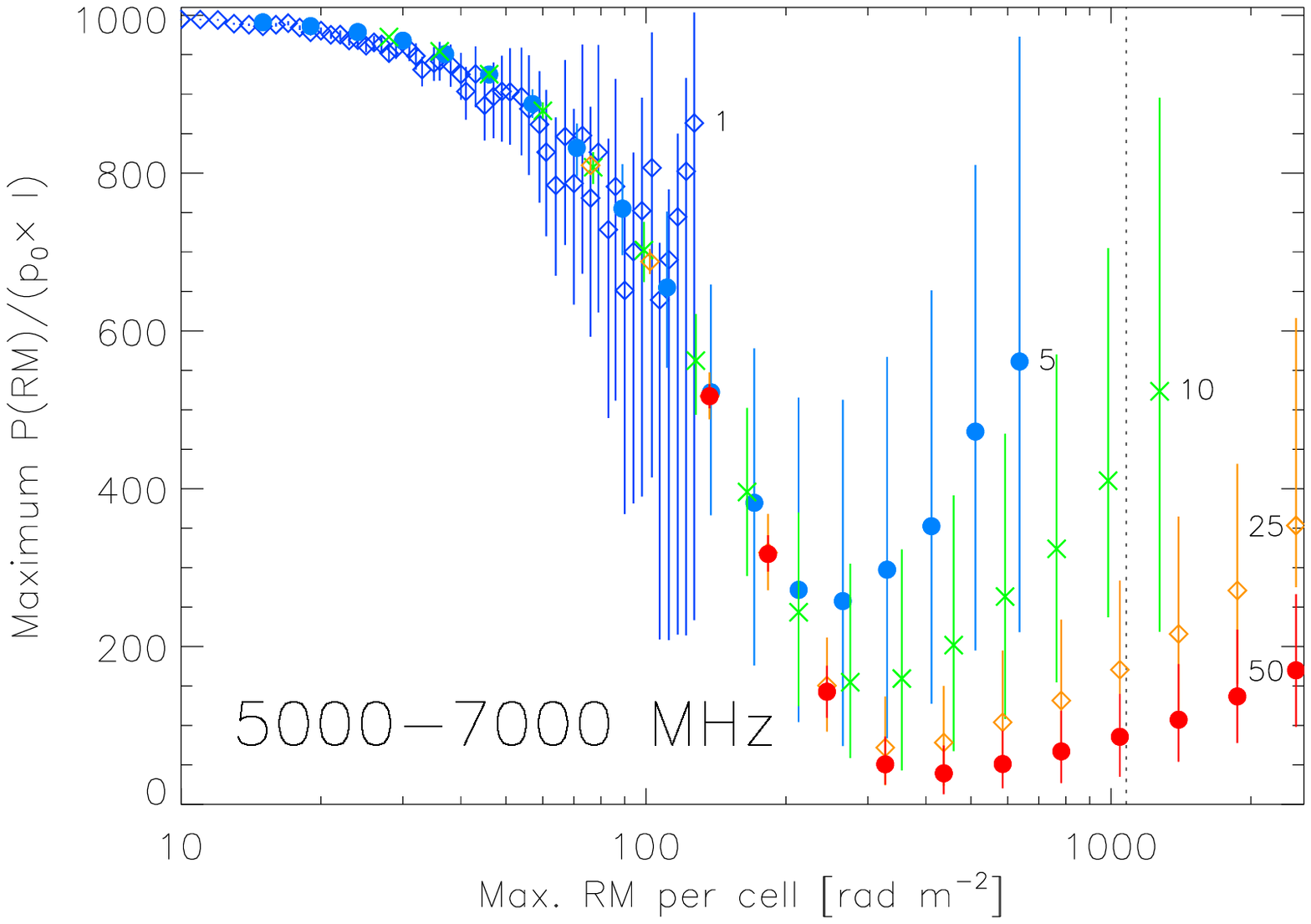}} 
    \vspace{-3ex}
  \end{minipage} 
\caption{The equivalent of Fig.~\ref{turb_monolayer.fig} when there are eight layers of turbulent cells. Note that the maximum RM of a single line of sight through the Faraday screen is eight times the maximum RM per cell which we plot along the x-axis of the panels.
}
\label{turb_quadlayer.fig}
\end{figure*}

\subsubsection{Geometries with turbulent magnetic fields}\label{rm_spectra_turbulent}
We investigate how RM spectra can help us understand the properties of the turbulent magnetic field that we simulated in \S~\ref{turbulent.sec} (its strength, coherence length, and the number of layers with turbulent cells) in three ways: first,  using the peak flux density in the RM spectrum, second, the shape of the RM spectrum, and third, the difference between the RM spectra of partial coverage models and our random walk models. 
Previously, \citet{bernet2012} investigated how RM synthesis can be used to investigate the properties of turbulent Faraday screens for a small number of turbulent cells and a Gaussian pdf(RM). We consider models where the number of sightlines through the turbulent foreground varies over a much wider range, and we consider different probability density functions.

First, we investigate how the peak normalized flux density in the RM spectrum depends on the coherence length and strength of the magnetic field. 
Equation~\ref{p_obs_turbulent} expresses how polarized emission from the background source is divided into $N_\mathrm{l.o.s.}$ polarization vectors, one vector for each line of sight through the turbulent foreground medium. 
These polarization vectors all have the same length, equal to $P/N_\mathrm{l.o.s.}$, but different RMs.
The frequency and RM spectra combine the behaviour of each of these polarization vectors; 
the RM spectrum consists of $N_\mathrm{l.o.s.}$ peaks spread over a range in RM between $N_\mathrm{layers}\times \mathrm{RM}_\mathrm{max}$, where $N_\mathrm{layers}$ indicates the number of layers of cells in the turbulent foreground screen, and $\mathrm{RM}_\mathrm{max}$ is defined by equation~\ref{rm_max}. 

In Figures~\ref{turb_monolayer.fig} and \ref{turb_quadlayer.fig} we show the peak polarized flux density in the RM spectrum for a number of magnetic field strengths, values of $\mathrm{RM}_\mathrm{max}$, and two different thicknesses of the Faraday-rotating screen. 
The background source is identical to the source that we simulated in \S~\ref{turbulent.sec}: it has a diameter of 25 parsec and illuminates the Faraday screen uniformly.
The electron density in the foreground screen is 10 cm$^{-3}$ and the magnetic field strength is 1, 5, 10, 25, or 50 $\mu$G.
We considered a Faraday screen that consists of eight layers because its pdf(RM) is very close to Gaussian (Appendix~\ref{Appendix_D}).
The four panels in these figures correspond to the four frequency windows that we used in \S~\ref{rm_spectra_turbulent}.  
For curves of a given colour or given symbols, the only variable is the coherence length of the magnetic field in the Faraday screen: the coherence length decreases when going from right to left in these panel as the number of sightlines $N_\mathrm{l.o.s.}$ increases.
The smallest number of sightlines that we considered is two since one sightline produces no depolarization. 
For magnetic fields of 25 and 50 $\mu$G two sightlines produce very large $\mathrm{RM}_\mathrm{max}$, therefore, instead of starting with two sightlines we started with that number of sightlines which produces $\mathrm{RM}_\mathrm{max}$ = 2500 rad~m$^{-2}$. 
We stop increasing the number of sightlines either when the simulation reached $\mathrm{RM}_\mathrm{max}$ = 10 rad~m$^{-2}$ or when the number of sightlines becomes larger than 5000. 

To understand the behaviour of the points shown in Figs~\ref{turb_monolayer.fig} and \ref{turb_quadlayer.fig} consider a single $B_\mathrm{turb}$ and let $N_\mathrm{l.o.s.}$ vary. When $N_\mathrm{l.o.s.}$ is small, RM$_\mathrm{max}$ ($\propto 1/\sqrt{N_\mathrm{l.o.s.}}$) is large, and since there are not many cells, the polarized flux density that is associated with each peak in the RM spectrum ($= 1000/N_\mathrm{l.o.s.}$) is large. 
The mean separation between two peaks in the RM spectrum is equal to
\begin{eqnarray}
\langle \Delta\mathrm{RM}\rangle = 
\frac{2\mathrm{RM}_\mathrm{max}}{N_\mathrm{l.o.s.}} \propto 
\frac{B_\mathrm{turb}N_\mathrm{layers}}{N_\mathrm{l.o.s.}^{3/2}}
\label{mean_rm}
\end{eqnarray}
When $N_\mathrm{l.o.s.}$ increases the length of the polarization vector associated with each sightline decreases, and the peaks move closer together in the RM spectrum because the RM range over which they are distributed ($|$RM$| \leq \mathrm{RM}_\mathrm{max}$) decreases. It becomes more likely that some of the peaks in the RM spectrum overlap, starting with a partial overlap of the RM spread functions. 
This explains why the peak polarized flux density in the RM spectrum does not decrease monotonically when $N_\mathrm{l.o.s.}$ is increased.
Increasing $N_\mathrm{l.o.s.}$ even further moves the peaks closer together, and as a result of overlapping RM spread functions the peak flux density in the RM spectrum will rise again. 
The intrinsic position angles of the emitted radio waves, $\chi_0$, are the same in our simulation, therefore the polarization vectors of the individual peaks will align almost perfectly once RM$_\mathrm{max}$ becomes much smaller than the FWHM of the RM spread function, which leads to very high peak polarized flux densities in the RM spectra.

From comparing the four panels in Figs~\ref{turb_monolayer.fig} and \ref{turb_quadlayer.fig}, and matching panels in both Figures, we draw the following conclusions. 
\begin{enumerate}

\item If $N_\mathrm{l.o.s.}$ is small (the field coherence length is large compared to the size of the background source) and the peaks in the RM spectrum do not overlap then models with different $B_\mathrm{turb}$ and $N_\mathrm{layers}$ produce the same peak polarized flux density in the RM spectrum, albeit at different RM$_\mathrm{max}$. When $N_\mathrm{l.o.s.}$ is small the RM spectrum is not sampled well, therefore it is difficult to determine either RM$_\mathrm{max}$ or the shape of pdf(RM) and from that $N_\mathrm{layers}$.

\item Increasing $B_\mathrm{turb}$ whilst keeping $N_\mathrm{layers}$ fixed increases the RM$_\mathrm{max}$ where the peak polarized flux density in Figs~\ref{turb_monolayer.fig} and \ref{turb_quadlayer.fig} is at its lowest. 
The minimum in these Figures depends on the mean separation between the peaks in the RM spectrum relative to the FWHM of the RM spread function. 
To keep the mean separation between peaks in the RM spectrum (equation~\ref{mean_rm})  fixed when $B_\mathrm{turb}$ is increased, $N_\mathrm{l.o.s.}$ only has to increase by a small amount.
Since RM$_\mathrm{max}$ $\propto B_\mathrm{turb}/\sqrt{N_\mathrm{l.o.s.}}$ increasing $B_\mathrm{turb}$ decreases the RM$_\mathrm{max}$ of the minimum in Figs~\ref{turb_monolayer.fig} and \ref{turb_quadlayer.fig}.

\item If $N_\mathrm{l.o.s.}$ is large, models with different $B_\mathrm{turb}$ produce the same peak flux density in the RM spectrum at the same RM$_\mathrm{max}$. 
It is then impossible to determine $B_\mathrm{turb}$ from the measured peak polarized flux density even if RM$_\mathrm{max}$ is known. 
However, if $N_\mathrm{l.o.s.}$ is large pdf(RM) is sampled well, and it might be possible to determine both RM$_\mathrm{max}$ and $N_\mathrm{layers}$ from the shape of pdf(RM).

\item Sources with a small field coherence length (such that $N_\mathrm{l.o.s.}$ is large) could be invisible in low-frequency windows and only appear in high-frequency windows. 
This is because the peak polarized flux density in the RM spectrum can be higher in high-frequency observing windows than in low-frequency windows.
In low-frequency windows, if $N_\mathrm{l.o.s.}$ is large the peaks in the RM spectrum have a small amplitude, and the peaks in the RM spectrum do not overlap.
At higher frequencies the RM spread functions are wider, and the peaks start to overlap in the RM spectrum.
Because we use the same intrinsic position angle $\chi_0$ for all polarization vectors, this overlap can increase the peak polarized flux density in the RM spectrum. 
This observation could lead to a trade-off between being able to detect a source (high peak polarized flux density in the RM spectrum caused by overlapping RM spread functions) versus being able to accurately measure the RMs of the individual peaks in the RM spectrum (which requires narrow RM spread functions).

\item If the RM spectrum is barely resolved and $N_\mathrm{l.o.s.}$ is small, as is the case for the 5-7 GHz window for $B_\mathrm{turb}$ = 1 $\mu$G, the peak polarized flux densities show complex behaviour.
\end{enumerate}

\begin{figure}
\resizebox{\hsize}{!}{\includegraphics{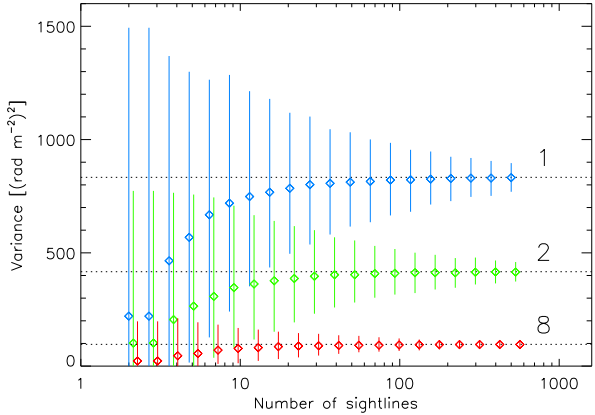}}
\caption{Variances calculated for a Faraday screen with various number of sightlines passing through them, for Faraday screens consisting of 1, 2, and 8 layers of turbulent cells. The maximum RM for a single turbulent cell is equal to 50 rad~m$^{-2}$ divided by the number of layers, which keeps the maximum total RM through the screen fixed and independent of the number of layers. Diamonds indicate the median variance, error bars indicate the most compact 95 per cent confidence interval, and the horizontal dotted lines indicate the variance that we calculated analytically for each of the Faraday screens.
}
\label{turb_n_los_comparison.fig}
\end{figure}

Second, if pdf(RM) is sampled sufficiently finely it should be possible to determine RM$_\mathrm{max}$ and the number of layers in the Faraday screen from the shape of the RM spectrum. 
We estimate the number of sightlines that is required to do this using a Monte Carlo simulation, drawing a number of sightlines from pdf(RM) for a Faraday screen that consists of 1, 2, or 8 layers of turbulent cells, and repeating this process 2000 times. 
The pdf(RM) of a Faraday screen which consists of eight layers of cells is almost identical to a Gaussian pdf(RM).
Fig.~\ref{turb_n_los_comparison.fig} shows the median variance, the 95 per cent confidence interval of the variances, and the analytically calculated variance for these three different Faraday screens. 
The range in RM that is spanned by pdf(RM) is the same for the three Faraday screens, only the shape of the distribution is different.
As can be expected, if the number of sightlines is small pdf(RM) is not well sampled, and the scatter in the variances for the different Faraday screens is large: it is impossible to identify screens of different thickness based on the shape of their RM spectrum.
Once there are at least several tens of sightlines it becomes possible to identify the shape of pdf(RM) and from that the number of layers in the Faraday screen.
In our simulation we did not include the finite width of the RM spread function. Because of this finite width the peaks in the RM spectrum from the individual sightlines will overlap, which complicates finding the shape of pdf(RM) of the Faraday screen.

Third, and final, we consider how RM spectra of turbulent foregrounds that we modelled in \S~\ref{multilayer.sec} differ from RM spectra of partial coverage models. In a partial coverage model the Faraday-rotating screen consists of a contribution by a Gaussian pdf(RM) and a contribution by the background source that does not undergo Faraday rotation (equation~\ref{pi_partial_coverage.eqn}). 
The latter contribution shows up in the RM spectrum as a peak at RM = 0 rad~m$^{-2}$, which combines with the Gaussian pdf(RM) from sightlines that pass through the Faraday screen. 
The relative heights of the Gaussian pdf(RM) and the single peak that is produced by sightlines that do not pass through the Faraday screen depends on the fraction of sightlines which pass through the Faraday screen.
For a source that emits uniformly across its surface, this fraction is equal to the fraction of the surface of the source that is covered by the Faraday screen.
Because instrumental leakage can show up as a strong signal at RM = 0 rad~m$^{-2}$, telescope leakages should be calibrated well to distinguish between the peak in polarized flux density at RM = 0 rad~m$^{-2}$ that is produced by the partial coverage model and the same peak that is produced by polarization leakage in the telescope.

\section{Summary}\label{s-conclusions}
We have modelled sources with large-scale and turbulent magnetic fields, and predicted frequency spectra for these sources between 200 MHz and 10 GHz, where we included the finite width of the frequency channels in our analysis.
In these models Faraday rotation takes place in front of the source of the emission.
We considered uniform and Gaussian sources on the sky with linear transverse RM gradients, and cylinders and spheroids with azimuthal magnetic fields.
The cylinders and spheroids can be inclined with respect to the plane of the sky, and for the spheroids we considered different axis ratios ranging from 1:1:0.25 (oblate), 1:1:1 (sphere), to 1:1:4 (prolate).
These source types with large-scale magnetic fields do not show net Faraday rotation: this is because integrating over the surface of the source combines sightlines with the same polarized flux density but RMs of opposite sign.
As a result any net RM of the source must be produced in the foreground, further away from the source.
We derived the probability density functions (pdfs) of RM for turbulent screens with between one and twenty layers of turbulent cells.
A single layer of cells has a uniform pdf(RM), while a Gaussian pdf(RM) is a good approximation for a Faraday screen that consists of more than about four layers of turbulent cells.
We show how increasing the thickness of the Faraday-rotating layer changes pdf(RM) and the properties of the polarized flux density spectra as a function of frequency and RM.

At high frequencies all the source types which we considered show a similar drop-off in polarized flux density with decreasing frequency.
Some source types show secondary maxima in their polarized flux density spectra at low and intermediate frequencies, which helps with their identification.

The model by \citet{burn1966} is often used to explain depolarization in turbulent foregrounds. 
It requires that many sightlines pass through the Faraday screen, which means that the field coherence length is small compared to the extent of the background source.
This large number of sightlines in Burn's model also produces complete depolarization at low and intermediate frequencies.
Partial coverage models were developed to explain why some sources are not completely depolarized at these frequencies.
We show that a Monte Carlo model of a small number of sightlines passing through a turbulent Faraday screen predicts a drop-off at high frequencies similar to the Burn depolarization model, and a plateau in polarized flux density at low and intermediate frequencies.
The polarized flux density of this plateau can be calculated from a random walk of the polarization vectors.
Spectra that are produced by the Monte Carlo model are very similar in shape to spectra produced by partial coverage models; therefore, Monte Carlo models can be considered as an alternative to partial coverage models.

We calculate RM spectra for four frequency windows: 350--900 and 950--1760 MHz, and 1.3--3.1 and 5.0--7.0 GHz.
At low frequencies, sources are strongly depolarized and might not be detected above the instrumental polarization level. However, with sufficient sensitivity RMs can be determined accurately.
Similar to the missing short-spacing problem in interferometry, if only low-frequency data are used the reconstructed RM spectrum of the uniform source shows a local minimum at its centre. 
Such sources can show up in RM spectra as two peaks above the instrumental polarization level.

Each sightline through a turbulent Faraday screen produces a single peak in the RM spectrum; the reconstructed RM spectrum is a superposition of these peaks convolved with the RM spread function.
We analyse the peak height and shape of the RM spectra when we change the strength and coherence length of the turbulent magnetic field and the number of layers with turbulent cells. 
Decreasing the field coherence length increases the number of sightlines and therefore the number of peaks in the RM spectrum. At the same time the polarized flux of the background source is distributed over more peaks, and the range in RM over which the peaks are distributed is reduced.
The combined effect of this is that the peak amplitude in the RM spectrum first decreases when the field coherence length is decreased, then increases when the individual peaks in the RM spectrum (convolved with the RM spread function) overlap.
When more then several tens of sightlines pass through a thin, turbulent foreground screen the shape of pdf(RM) can be determined reasonably accurately, and with this information the number of layers of turbulent cells can be derived.
Finally, in a partial coverage model the part of a source that is not covered by a turbulent screen produces a peak in the RM spectrum at 0 rad~m$^{-2}$.
Random walk models do not show such a peak; therefore RM spectra can tell which of the two models applies.

The modelling framework we provide can be extended to any frequency coverage, frequency range, and channel width of the observations, and can be used to explore more complex source types and combinations of source types.

\section*{Acknowledgements}
We would like to thank Alex Kraus (Max Planck Institute for Radio Astronomy) and Cathy Horellou (Chalmers University of Technology) for their comments which helped improve the manuscript.
KJL gratefully acknowledges support from the National Natural Science Foundation of China (Grant no.~11373011).

\appendix
\section{Frequency spectra of a cylinder with an azimuthal magnetic field}\label{Appendix_A}
In this Appendix we model the frequency spectrum of an emitting cylinder with a magnetic field that wraps around the circular cross-section of the cylinder. Faraday rotation occurs in a boundary layer between the inner, emitting cylinder and a coaxial outer cylinder. 
The magnetic field inside the emitting cylinder $\bmath{B}_\mathrm{em}$ has a constant strength, and points along the major axis of that cylinder. 
The inner cylinder has a radius $R$, while the outer cylinder has a radius $R'$. 
We define a Cartesian coordinate system $(x,y,z)$ where $x$ points towards the observer, $y$ lies along the projection of the short axis of the cylinder on the sky, and $z$ along the projection of the long axis. 
We will use $\theta$ for the angle between the plane of the sky and the major axis of the emitting cylinder. 
The observed monochromatic net polarization vector is given by equation~\ref{p_monochromatic_source}.
The $E$ vector of the emitted polarization vector $\bmath{P}_\mathrm{em}$ lies along the $y$ axis of the coordinate system that we chose. 
We define the intrinsic position angle $\chi_0$ to be equal to zero degrees if $\bmath{P}_\mathrm{em}$ points in the direction of $\hat{y}$.

Because in our model the synchrotron-emitting and Faraday-rotating layers are not mixed we can solve equation~\ref{p_monochromatic_source} by first calculating the emitted polarized flux density along each line of sight, then calculating the amount of Faraday rotation of this emission.
To simplify our analysis we will not consider sightlines that pass through the polar caps of the cylinder.

The monochromatic volume emissivity for synchrotron radiation, $\epsilon_\nu$, depends on the strength of the magnetic field inside the cylinder and the inclination of the cylinder relative to the sky $\theta$ as $\epsilon_\nu = \mathrm{K} \left|\bmath{B}_\mathrm{em}\cos{\theta}\right|^{\alpha+1}\nu^{-\alpha}$ (e.g., \citealt{rybicki1979}). `K' depends on the mass and charge of the synchrotron emitting particles and on the spectral index of the synchrotron emission $\alpha$ ($S_\nu \propto \nu^{-\alpha}$). 
The flux density of a single line of sight is found by integrating the volume emissivity along the line of sight:
\begin{eqnarray}
\bmath{P}_\mathrm{em}\left(y,z\right) = \int \epsilon_\nu\left(x,y,z\right) \sqrt{1-\left(y/R\right)^2}/\left|\cos{\theta}\right| \mathrm{d}x\, ,
\nonumber
\end{eqnarray}
where the factor $\sqrt{1-\left(y/R\right)^2}$ corrects for the angle between the line of sight and the near surface of the cylinder, and $1/\left|\cos{\theta}\right|$ corrects for the inclination of the cylinder with respect to the plane of the sky.

In our model the frequency dependence of the emissivity $\nu^{-\alpha}$ is absorbed into the product of the Stokes $I$ spectrum and the wavelength-independent polarization fraction $p_0$; by requiring that the source emits 1000 units of polarized flux density independent of $\theta$ the entire factor $\mathrm{K} \left|\bmath{B}_\mathrm{em}\cos{\theta}\right|^{\alpha+1}/\left|\cos{\theta}\right|$ is fixed.
Therefore the volume emissivity is constant throughout the emitting cylinder, and the polarized flux density of a single line of sight depends only on the geometry of the emitting cylinder with respect to the line of sight:
\begin{eqnarray}
\frac{\bmath{P}_\mathrm{em}\left(y,z\right)}{p_0 I} \propto  1-\left(\frac{y}{R}\right)^2\, . 
\nonumber
\end{eqnarray}

The amount of Faraday rotation of a single sightline can be calculated by integrating the product of the magnetic field strength along the line of sight and the local free electron density in the boundary layer between the inner and outer cylinder:
\begin{eqnarray}
\mathrm{RM}\left(y,z\right) & = & \int_{\sqrt{R^2-y^2}}^{\sqrt{\left(R'\right)^2-y^2}} 0.81 n_\mathrm{e}|\bmath{B}| \frac{y}{\sqrt{x^2+y^2}}\mathrm{d}x \nonumber \\
& = & 0.81 n_\mathrm{e}|\bmath{B}| y \ln\left(\frac{R' + \sqrt{\left(R'\right)^2 - y^2}}{R + \sqrt{R^2 - y^2}}\right)\, ,
\label{cylinder_RM}
\end{eqnarray}
where the boundary layer has an electron density $n_\mathrm{e}$ and is threaded by a magnetic field of amplitude $|\bmath{B}|$. If the source is inclined relative to the plane of the sky then the pathlength through the Faraday screen increases as $1/\left|\cos{\theta}\right|$, which exactly cancels the projection of the magnetic field along the line of sight in the Faraday-rotating layer ($\propto \left|\cos{\theta}\right|$). Therefore RM$\left(y,z\right)$ does not depend on the inclination of the source relative to the sky.

The maximum RM, RM$_\mathrm{max}$, is found at $y=R$, and equation~\ref{cylinder_RM} can be expressed in terms of RM$_\mathrm{max}$ and the geometry of the system as 
\begin{eqnarray}
\frac{\mathrm{RM}\left(y,z\right)}{\mathrm{RM}_\mathrm{max}} & = & \frac{y}{R} 
\frac{\ln\left(\frac{R' + \sqrt{\left(R'\right)^2 - y^2}}{R + \sqrt{R^2 - y^2}}\right)}
{\ln\left( \frac{R'}{R} + \sqrt{ \left(\frac{R'}{R}\right)^2 - 1}\right)}\, .
\label{cylinder_RM_max}
\end{eqnarray}
If $R'$ is only slightly larger than $R$ equation~\ref{cylinder_RM_max} gives almost identical results as when we replace the line-of-sight component of the magnetic field, which varies in amplitude along the line of sight, with the value halfway through the Faraday-rotating boundary layer:
\begin{eqnarray}
\frac{\mathrm{RM}\left(y,z\right)}{\mathrm{RM}_\mathrm{max}} & = & 
\frac{y}{R} \frac{\left(\sqrt{\left(R'\right)^2 - y^2} - \sqrt{R^2 - y^2} \right)}{\left(\sqrt{\left(R'\right)^2-R^2} \right)}\, .
\label{cylinder_thin_boundary_approximation}
\end{eqnarray}

\begin{figure}
\resizebox{\hsize}{!}{\includegraphics{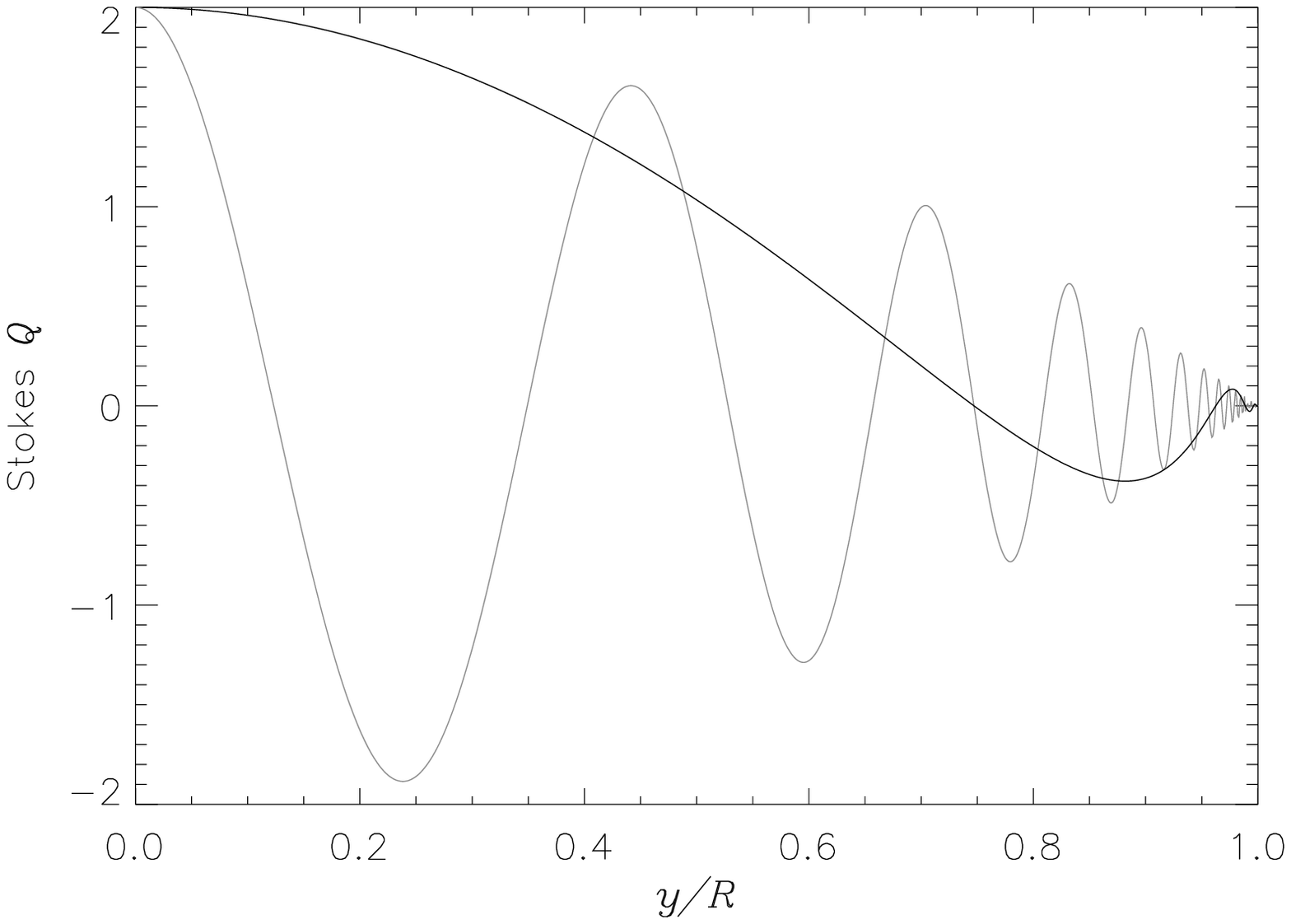}}
\caption{The behaviour of Stokes $Q$ (with arbitrary units) with position along the minor axis of the cylinder, for RM$_\mathrm{max}$ = 1000 rad~m$^{-2}$ and frequencies of 3.0 GHz (dark grey) and 1.0 GHz (light grey). The Faraday-rotating boundary layer has a thickness 1/10th of the radius of the synchrotron-emitting cylinder.  
This profile does not depend on $\theta$ because of the normalization we used.
}
\label{p_profile_cylinder.fig}
\end{figure}

In Fig.~\ref{p_profile_cylinder.fig} we show the observed values of Stokes $Q$ for RM$_\mathrm{max}$ = 1000 rad~m$^{-2}$ as a function of $y$ for frequencies of 1.0 and 3.0 GHz. 
Our definition for $\chi_0$ = 0\degr\ results in the emitted polarization vector $\bmath{P}_\mathrm{em}$ showing up only in Stokes $Q$, Stokes $U$ is zero.
The cylinder that we simulated does not show any net Faraday rotation: the symmetry of the system is such that points on opposite sides of the major axis of the cylinder (at $\pm y$) emit the same polarized flux density and are Faraday rotated in opposite directions, which cancels Faraday rotation for the source as a whole.
Therefore all net polarization vectors for the source as a whole will show up only in Stokes $Q$, independent of the observing frequency.

To calculate the monochromatic polarization vector for the source as a whole we integrated over the minor axis $y$ of the cylinder using Romberg's method. 
To integrate across the finite width of the individual frequency channels, and to correctly include changes in the orientation of the monochromatic net polarization vector within each frequency channel, we divide each channel into subintervals that are narrow enough so that we can calculate the net polarization of each subinterval using the trapezium rule.
We use forty subintervals for each $2\upi$ revolution of the position angles that is induced by the largest RM of the geometry that we modelled, RM$_\mathrm{max}$. 
By combining the net polarization vectors of the subintervals we calculate the net polarization vector of a single frequency channel as a whole.
We only include a frequency channel in our analysis if the length of the polarization vector of the first subinterval is more than ten times the numerical accuracy $\epsilon$ defined by equation~\ref{epsilon_definition}; otherwise we stopped calculating the frequency spectrum altogether.
The first subinterval lies at the low-frequency-end of each frequency channel.

\section{Frequency spectra of an ellipsoid with an azimuthal magnetic field}\label{Appendix_C}
\begin{figure}
\resizebox{\hsize}{!}{\includegraphics{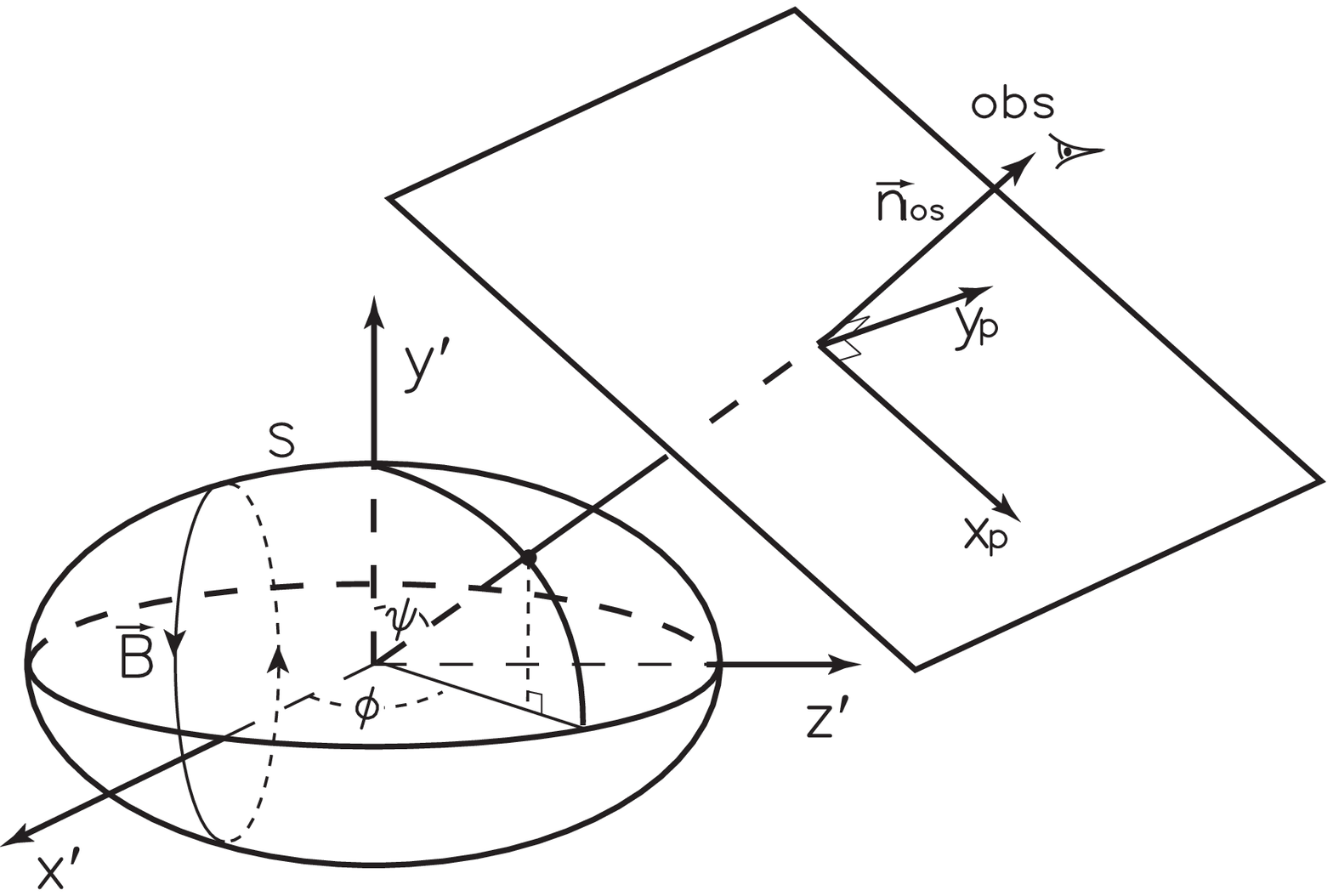}}
\caption{Geometry relating the coordinate system of the ellipsoid with surface S and magnetic field $\bmath{B}$ to the line of sight `los' towards the observer `obs' .The orthonormal vectors $\bmath{x}_\mathrm{p}$ and $\bmath{y}_\mathrm{p}$ lie in the plane of the sky.
}
\label{ellipsoid_geometry.fig}
\end{figure}

In this Appendix we model the frequency spectrum of a synchrotron-emitting ellipsoid. 
This geometry consists of two nested ellipsoids with a shared coordinate system, shown in Fig.~\ref{ellipsoid_geometry.fig}.
The boundary of the inner, synchrotron-emitting ellipsoid is described by $(x'/A)^2\ +\ (y'/B)^2\ +\ (z'/C)^2\ =\ 1$, and the magnetic field inside the emitting ellipsoid points along the $z'$ axis.
Faraday rotation occurs in a layer between the inner ellipsoid and an outer ellipsoid whose surface is given by $(x'/A_{\rm B})^2+(y'/B_{\rm B})^2+(z'/C_{\rm B})^2=1$. 
The magnetic field in the Faraday-rotating layer wraps around the major axis ($z'$ axis) of the inner ellipsoid and is parallel to the surface of the inner ellipsoid.
The thickness of the Faraday-rotating boundary layer varies because the Faraday-rotating layer is nested between two ellipsoids; by comparison, in our model of the emitting cylinder (Appendix~\ref{Appendix_A}) the thickness of the Faraday-rotating boundary layer was the same everywhere.

In the $(x',y',z')$ coordinate system of the ellipsoid the line of sight towards the observer, indicated by the unit vector $\bmath{n}_\mathrm{los}$, is specified by the angles $\phi$ and $\psi$, where
\begin{eqnarray}
	{\bf n}_{\rm los}=\{ \cos\psi ,\sin\psi\cos\phi, \sin\psi\sin\phi\}\, , \nonumber
\end{eqnarray}
We choose the two mutually perpendicular unit vectors $\bmath{x}_{\rm p}$ and $\bmath{y}_{\rm p}$ in the plane of the sky as the coordinate basis, where
\begin{eqnarray}
	{\bf x}_{\rm p}&=&\{ -\sin\psi,  \cos\phi\cos\psi, \sin\phi\cos\psi, 	
	\}\, , \nonumber\\
	{\bf y}_{\rm p}&=&\{0, -\sin\phi, \cos\phi \}\, . \nonumber
\end{eqnarray}
In the main text we will use $\phi = 90^\circ$. In that case $\psi = \theta$, the inclination of the major axis of the ellipsoid with respect to the plane of the sky.

To calculate the monochromatic net polarization vector $\bmath{P}\left(\nu\right)$, which is integrated over the surface of the source, we divide the surface of the source into a very fine grid of sightlines and calculate the polarization vector of each of these sightlines numerically.
Similar to the polarized emission from the cylinder that we considered in Appendix~\ref{Appendix_A} we assume that both the inclination angle of the magnetic field and the volume emissivity $\epsilon_\nu$ are uniform inside the emitting ellipsoid.
The inclination angle of the magnetic field can then be absorbed into the normalization of the monochromatic net polarization vector, and the emitted polarized flux density is simply proportional to the length of the line of sight through the emitting ellipsoid.
We calculate the polarized flux density of each sightline analytically; the proportionality constant is fixed by our requirement that the source emits 1000 units of polarized flux density.

To calculate the amount of Faraday rotation of each sightline we integrate equation~\ref{rm_definition} numerically through the Faraday-rotating boundary layer using the trapezium rule, with 40 sampling points per sightline.
Then we calculate the monochromatic net polarization vector $\bmath{P}\left(\nu\right)$ by adding the polarization vectors for a square grid of 1200$\times$1200 sightlines across the surface of the source projected onto the sky. 
For a few of our models we varied the number of sampling points through the Faraday rotating layer (20 or 40 points) or the density of the grid of sightlines across the surface of the source (800$\times$800 or 1200$\times$1200 sightlines). By comparing the results from these models we found that the fractional accuracy in the length of the polarization vector is better than 1\% of the emitted 1000 units of polarized flux density.
Close to the major axis of the source, where most of the polarized flux is generated, each $2\upi$ wrap of the position angles across the surface of the source is sampled with at least 2 sightlines.

To integrate $\bmath{P}\left(\nu\right)$ across the finite width of the frequency channels we calculate $\bmath{P}\left(\nu\right)$ at forty points for each 2$\upi$ wrap in position angle that is induced by the sightline with the largest RM, and we integrate across each frequency channel using the trapezium rule.

\section{The probability density function of RM of a turbulent foreground medium}\label{Appendix_D}
If there is only one layer of turbulent cells in front of the emission region then the RMs from this layer follow a uniform distribution. 
This can be shown as follows. We will use `pdf' as shorthand for the probability density function, and $i$ indicates the angle between the line of sight and the magnetic field. 
The observed RM = RM$_\mathrm{max}\cos{i}$, where RM$_\mathrm{max}$ is defined in equation~\ref{rm_max}.
Because probability mass is conserved, 
\begin{eqnarray}
\mathrm{pdf(RM)} = \frac{\mathrm{pdf}\left(i\right)}{\left|\mathrm{dRM/d}i\right|} = \frac{1/2 \left|\sin{i}\right|}{\mathrm{RM}_\mathrm{max}\left|\sin{i}\right|} = \frac{1}{2\mathrm{RM}_\mathrm{max}}\, ,
\nonumber
\end{eqnarray}
therefore pdf(RM) is uniform between $\pm\ \mathrm{RM}_\mathrm{max}$.
$\mathrm{pdf}\left(i\right)$ is equal to the fraction of the surface on the unit sphere between $i$ and $i+\mathrm{d}i$.

For two layers of turbulent cells the pdf for measuring $\mathrm{RM}_{1,2}\ =\ \mathrm{RM}_1\ +\ \mathrm{RM}_2$, $\mathrm{pdf}\left(\mathrm{RM}_{1,2}\right)$, can be found as the convolution $\mathrm{pdf}\left(\mathrm{RM}\right)\ast \mathrm{pdf}\left(\mathrm{RM}\right)$, 
\begin{eqnarray}
\mathrm{pdf}\left(\mathrm{RM}_{1,2}\right)\ =\ \int_{-\infty}^{\infty}\mathrm{pdf}\left(\mathrm{RM}_{1,2} - \mathrm{RM}\right)\mathrm{pdf}\left(\mathrm{RM}\right)\mathrm{dRM}\, ,
\end{eqnarray}
\noindent
which is triangular. For three layers, $\mathrm{pdf}\left(\mathrm{RM}_{1,2,3}\right)\ =\ \left\{\mathrm{pdf}\left(\mathrm{RM}\right)\ast\mathrm{pdf}\left(\mathrm{RM}\right)\right\}\ast
\mathrm{pdf}\left(\mathrm{RM}\right)\ =\ \mathrm{pdf}\left(\mathrm{RM}_{1,2}\right)\ast\mathrm{pdf}\left(\mathrm{RM}\right)$. This way $\mathrm{pdf}\left(\mathrm{RM}_{1,..,\mathrm{N}}\right)$ can be calculated for any number N of layers with turbulent cells. 

\begin{figure}
\resizebox{\hsize}{!}{\includegraphics{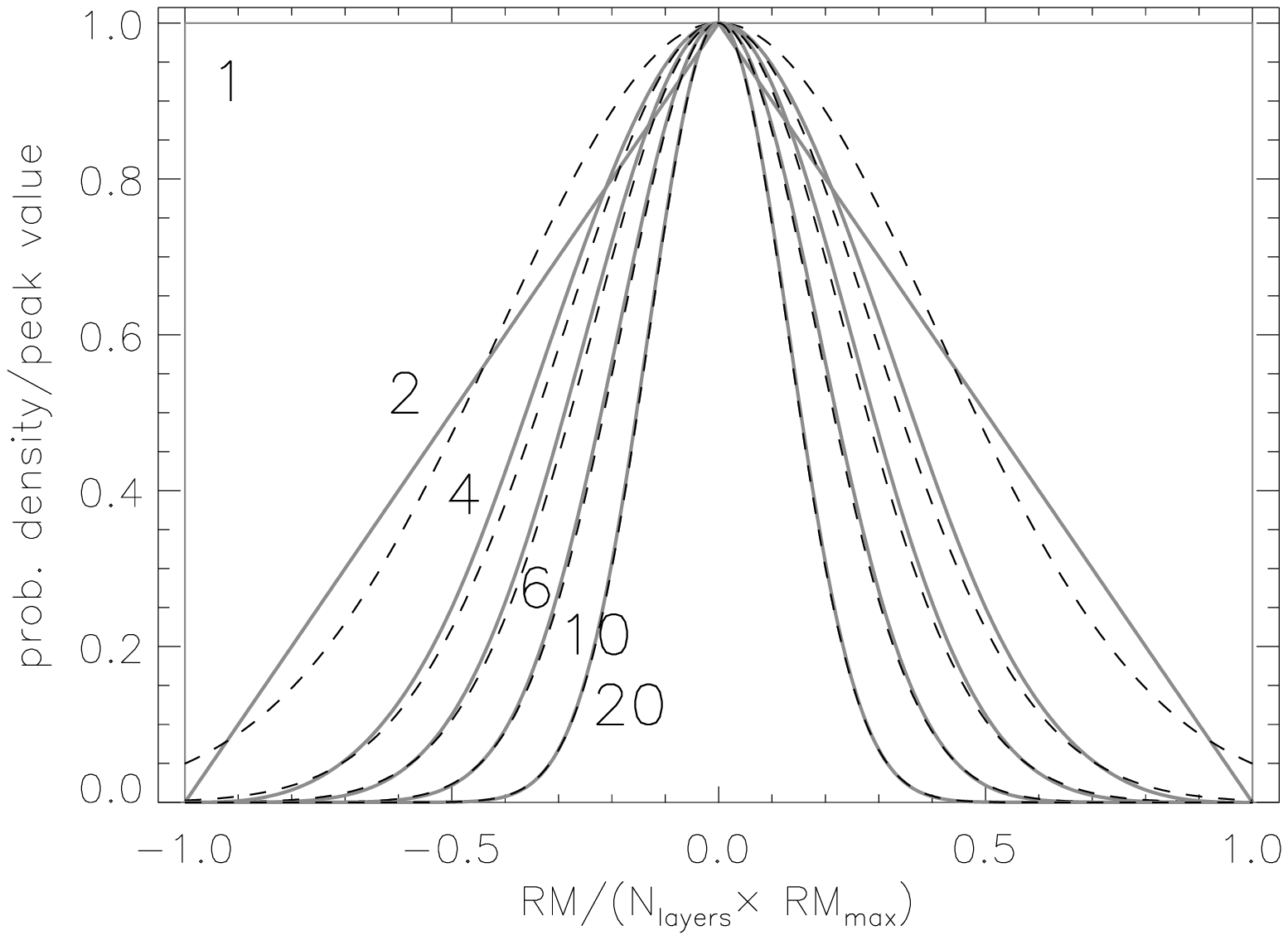}}
\caption{Probability density functions for one, two, four, six, ten, and twenty layers of turbulent cells (grey lines) and the normal distributions found by calculating the standard deviation of the probability density functions (dashed lines; see the text for details). 
}
\label{RM_pdf.fig}
\end{figure}

\begin{figure}
\resizebox{\hsize}{!}{\includegraphics{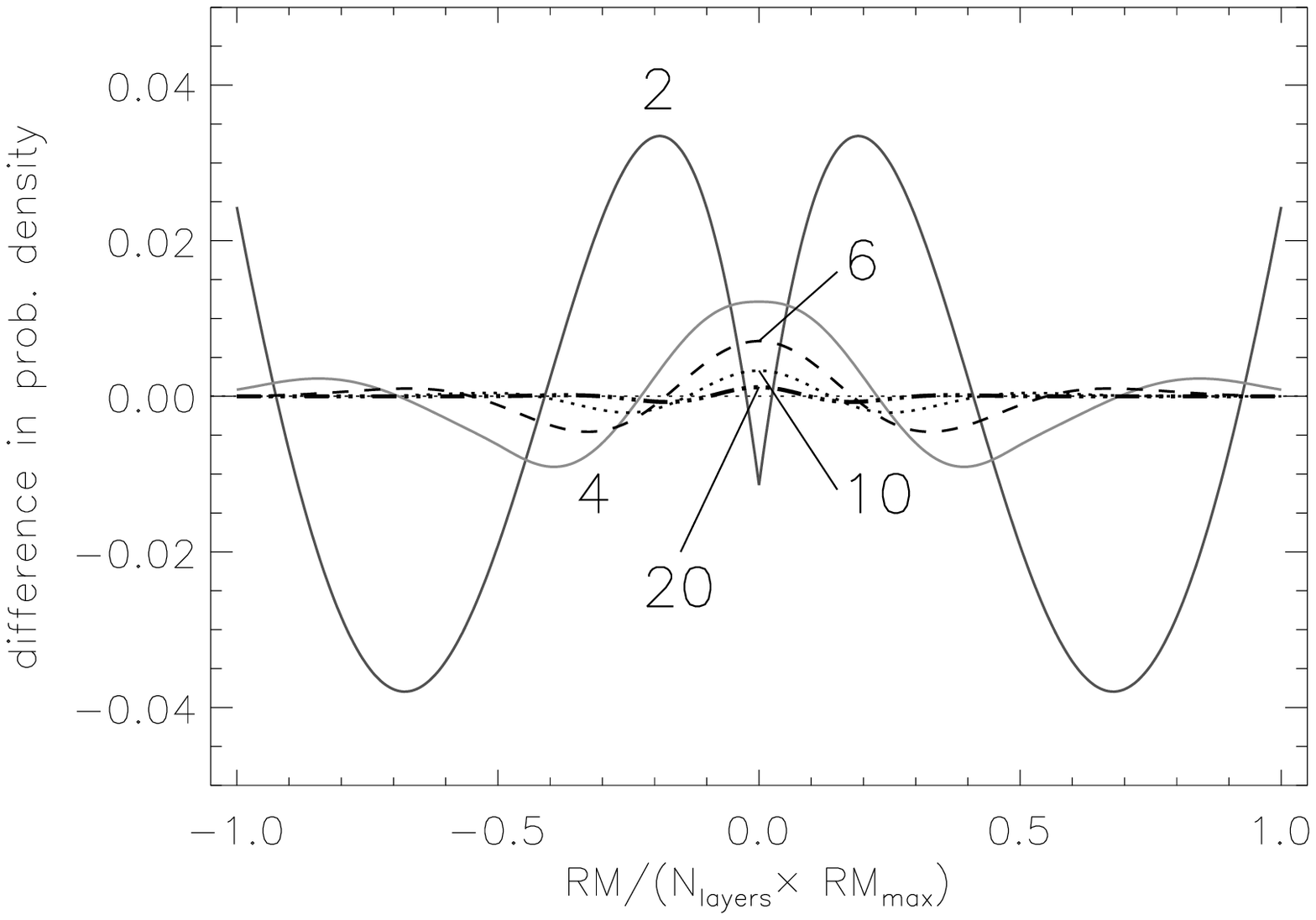}}
\caption{Difference in probability density between scaled normal distributions from Fig.~\ref{RM_pdf.fig} and the pdf(RM), for different numbers of layers with turbulent Faraday-rotating cells.
}
\label{RM_pdf_difference.fig}
\end{figure}

Fig.~\ref{RM_pdf.fig} shows pdfs for between one and twenty layers of turbulent, Faraday-rotating cells and their matching normal distributions, while Fig.~\ref{RM_pdf_difference.fig} shows the difference in probability density between the normal distributions and the pdf(RM). The standard deviation $\sigma$ of each normal distribution follows from the variance of probability density function that we calculated analytically.
The maximum difference between the normal distribution and pdf(RM) decreases from 0.0380 for 2 layers of Faraday-rotating cells to 0.0071 for six layers, 0.0033 for ten layers,  and 0.0012 for twenty layers of cells (these are the values for the pdfs themselves, not for pdf/max(pdf) that we show in Fig.~\ref{RM_pdf.fig}).

\end{document}